\documentclass[aps,prd,twocolumn,a4paper,superscriptaddress,10pt]{revtex4-1}
\usepackage{graphicx}
\usepackage{multirow}
\usepackage{latexsym}
\usepackage{hyperref}
\usepackage[british]{babel}
\usepackage{amsmath}
\usepackage{amsmath}
\usepackage{xcolor}
\begin{document}

\title{Quantifying the effect of cooled initial conditions on cosmic string network evolution}
\author{J. R. C. C. C. Correia}
\email{Jose.Correia@astro.up.pt}
\affiliation{Centro de Astrof\'{\i}sica da Universidade do Porto, Rua das Estrelas, 4150-762 Porto, Portugal}
\affiliation{Instituto de Astrof\'{\i}sica e Ci\^encias do Espa\c co, Universidade do Porto, Rua das Estrelas, 4150-762 Porto, Portugal}
\affiliation{Faculdade de Ci\^encias, Universidade do Porto, Rua do Campo Alegre 687, 4169-007 Porto, Portugal}
\author{C. J. A. P. Martins}
\email{Carlos.Martins@astro.up.pt}
\affiliation{Centro de Astrof\'{\i}sica da Universidade do Porto, Rua das Estrelas, 4150-762 Porto, Portugal}
\affiliation{Instituto de Astrof\'{\i}sica e Ci\^encias do Espa\c co, Universidade do Porto, Rua das Estrelas, 4150-762 Porto, Portugal}

\date{14 May 2020}

\begin{abstract}
Quantitative studies of the evolution and cosmological consequences of networks of cosmic strings (or other topological defects) require a combination of numerical simulations and analytic modeling with the velocity-dependent one-scale (VOS) model. In previous work, we demonstrated that a GPU-accelerated code for local Abelian-Higgs string networks enables a statistical separation of key dynamical processes affecting the evolution of the string networks and thus a precise calibration of the VOS model. Here we further exploit this code in a detailed study of two important aspects connecting the simulations with the VOS model. First, we study the sensitivity of the model calibration to the presence (or absence) of thermal oscillations due to high gradients in the initial conditions. This is relevant since in some Abelian-Higgs simulations described in the literature a period of artificial (unphysical) dissipation---usually known as cooling---is introduced with the goal of suppressing these oscillations and accelerating the convergence to scaling. We show that a small amount of cooling has no statistically significant impact on the VOS model calibration, while a longer dissipation period does have a noticeable effect. Second, in doing this analysis we also introduce an improved Markov Chain Monte Carlo based pipeline for calibrating the VOS model, Comparison to our previous bootstrap based pipeline shows that the latter accurately determined the best-fit values of the VOS model parameter, but underestimated the uncertainties in some of the parameters. Overall, our analysis shows that the calibration pipeline is robust and can be applied to future much larger field theory simulations. 
\end{abstract}
\maketitle
\allowdisplaybreaks
\section{\label{intr}Introduction}

Topological defects arise as consequences of symmetry breaking phase transitions in the early universe, by means of the Kibble mechanism \cite{Kibble:1976sj}. They arise naturally in many extensions of the Standard Model of particle physics \cite{Jeannerot:2003qv} and even in string theory \cite{Sarangi:2002yt}, thus being a natural fingerprint of these theories in the early Universe. The safest type of topological defect (in the pragmatic sense that they are not expected to overclose the Universe) are cosmic strings, which are one-dimensional tube-like objects. Given their ubiquitous nature they can be constrained by astrophysical observations \cite{LIGODefects,PlanckDefects} and are also a prime target for future facilities such as CORE \cite{CORE} or LISA \cite{LISA,Auclair:2019wcv}.

The canonical model of defect network evolution is the Velocity dependent One-Scale (VOS) model of Martins and Shellard  \cite{Martins:1996jp,Book}. For the case of cosmic strings, this was originally shown to successfully model the evolution of Nambu-Goto networks and Abelian-Higgs networks \cite{Moore:2001px}. More recently, taking advantage of progress in high-performance computing facilities \cite{Correia:2018gew}, we have shown that in order to obtain a more accurate description of the velocity dependencies of all model parameters one should use an extended version of this model \cite{Correia:2019bdl}. In the case of Abelian-Higgs strings, this required the generalization of the string momentum (or curvature) parameter and the introduction of an explicit scalar and gauge radiation energy loss term. These previous studies with Abelian-Higgs string simulations also demonstrated the feasibility of distinguishing (at least in a statistical sense) the effects of energy losses due to loop production and radiation in the evolution of string networks.

In the present work, we further exploit our GPU-accelerated Abelian-Higgs code \cite{Correia:2018gew} to continue the quantitative study of string network evolution, focusing on the assessment of the sensitivity of the model calibration to the presence (or absence) of thermal oscillations that arise from high gradients in the initial conditions. This is clearly an important question in principle, but it is also important on practical grounds, when comparing previous results in the literature. Indeed, in some previous Abelian-Higgs simulations \cite{Moore:2001px,Bevis:2006mj,Hindmarsh:2017qff}, since one is mainly interested in the properties of the string network once it has reached the scaling regime, a period of artificial (unphysical) dissipation is introduced with the goal of accelerating the convergence of the simulation to the expected scaling. This gradient flow period, colloquially known as a cooling phase, has the additional side effect of suppressing not only these early thermal oscillations but also subsequent radiation losses. On the other hand, in our previous work \cite{Correia:2018gew,Correia:2019bdl} no such cooling period was used, one reason for this being that the VOS model enables the explicit modelling of this radiation. A legitimate question therefore arises as to whether the two types of early evolutions for the numerical simulations lead to results for the network properties (e.g., defect densities and average velocities) that are directly comparable. A related question is how any such differences may impact the VOS model calibration. The primary goal of this work is therefore to provide an answer these questions.

The outline of the rest of work is as follows. We start in Sect. \ref{setup} with a succint description of our numerical simulation code, specifically introducing the three scenarios with different degrees of cooling that have been simulated. In Sect. \ref{compar}, after an equally brief introduction to the VOS model, we present and compare the model calibrations obtained from the three different cooling scenarios. Since the comparison between calibrations relies in part in accurately estimating the uncertainties in the model parameters, in Sect. \ref{calibs} we take the opportunity to go beyond previous calibration analyses (which were based on bootstrap methods) by introducing Markov Chain Monte Carlo (MCMC) based methods for calibrating the VOS model. Finally, our conclusions and a short outlook discussion can be found in Sect. \ref{conc}.

\section{\label{setup}Simulation setup and methodology}

There are two possible methods for simulating cosmic string networks in an expanding background: either by considering the infinitely thin string approximated by the Nambu-Goto action \cite{Blanco,VVO,BB,AS} or by using a field theory discretized on a lattice \cite{Moore:2001px,Bevis:2006mj,Correia:2019bdl,Hindmarsh:2017qff,Klaer:2019fxc}. While these two methods fully agree on the result that the network should achieve a scaling regime (where the mean string length scales linearly with time and the velocity remains constant) they disagree about the specific properties of the regime itself, most notably about which energy loss mechanism is responsible for sustaining this behavior. Naturally, this has direct implications for observational searches for cosmic strings, as well as for constraints on the underlying models.

Here we focus on the latter simulation method, following our recent work \cite{Correia:2018gew,Correia:2019bdl}, to which we refer the reader for a more detailed discussion. We choose a Lagrangian density describing a $U(1)$ locally invariant theory, where the breaking of the underlying symmetry supports the existence of a defect (i.e. the vacuum manifold is homotopically non-trivial). Such density can be written as,
\begin{equation}
	\mathcal{L}=|D_\mu \phi|^2 - \frac{\lambda}{4}(|\phi|^2 -1)^2 - \frac{1}{4e^2}F^{\mu \nu}F_{\mu \nu}\,,
\end{equation}
where $\phi$ is a complex scalar field, $D_\mu \phi$ is the gauge covariant derivative given by $D_\mu = \partial_\mu -iA_\mu$  $A_\mu$ is the gauge field (the gauge coupling $e$ has been absorbed), $A_\mu$ is the gauge field, the electromagnetic field tensor is given by $F_{\mu \nu} = \partial_\mu A_\nu - \partial_\nu A_\mu$, and $\lambda$ and $e$ are coupling constants (for which we respectively assume the values $2$ and $1$, corresponding to critical strings). In the temporal gauge ($A_0 =0$) in an expanding background ($g_{\mu \nu} = a^2 diag(-1,1,1,1)$), the equations of motion are,
\begin{equation}
\ddot{\phi} + 2\frac{\dot{a}}{a}\dot{\phi} = D^jD_j\phi -\frac{a^{2}\lambda}{2} (|\phi|^2 - 1) 
\end{equation}
\begin{equation}
\dot{F}_{0j} = \partial_j F_{ij} -2a^2 e^2 Im[\phi^* D_j \phi]\,.
\end{equation}

In order to avoid the problem of the string width becoming smaller than the comoving lattice as the simulation evolves, we apply the Press-Ryden-Spergel (PRS) prescription \cite{PRS,Bevis:2006mj} of modifying the equations of motion such that the comoving width of the strings is constant. This implies the following modified equations of motion:
\begin{equation}
\ddot{\phi} + 2\frac{\dot{a}}{a}\dot{\phi} = D^jD_j\phi -\frac{\lambda_0}{2} (|\phi|^2 - 1) 
\end{equation}
\begin{equation}
F_{0j} = \partial_j F_{ij} -e_0^2 Im[\phi^* D_j \phi]\,.
\end{equation}
Note that this implies the original coupling constants now have a dependency on time
\begin{equation}
e = e_0 a^{-1}
\end{equation}
\begin{equation}
\lambda = \lambda_0 a^{-2}\,,
\end{equation}
where $e_0$ and $\lambda_0$ can be thought of as the physical coupling parameters. For the explicit form of the discretized equations of motion, we refer the reader to \cite{Correia:2019bdl}.

Our choice of simulation parameters coincides with the one made in our previous work \cite{Correia:2019bdl}, specifically with lattice spacing $\Delta x = 0.5$, timestep size $\Delta t = 0.1$, box size of $512^3$ and all simulations being evolved until the horizon reaches half the box size.

From here we need only extract the necessary information about the networks which allows us to calibrate the VOS model. The quantities we are interested in are the mean velocity, denoted $v=\sqrt{\langle v^2\rangle}$, and the mean string separation in comoving coordinates, denoted $\xi$. In order to extract the velocity we use the velocity estimator from \cite{Hindmarsh:2017qff} which uses the equation of state parameter from the string (calculated with pressure and hamiltonian density weighted by the Lagrangian),
\begin{align}\label{diagvel}
  \langle v^2\rangle_{\omega} = \frac{1}{2} \bigg( 1+3\frac{\sum_x p_x \mathcal{L}_x}{\sum_x \rho_x  \mathcal{L}_x} \bigg)\,,
\end{align}
where $\mathcal{L}_x$ is the Lagrangian computed at some lattice site x. For computing the mean string separation we use the lattice discretized version of the winding from \cite{Kajantie:1998bg}, 
\begin{align}\label{diagxi}
  \xi_W = \sqrt{\frac{\mathcal{V}}{\sum_{ij,x} W_{ij,x}}}\,.
\end{align}
which identifies which lattice cells are pierced by strings: a cell face pierced by a string has $W_{i,j} \neq 0$. Again, for more details we refer the reader to \cite{Correia:2019bdl}, where these diagnostics have also been compared to alternative ones.

In order to apply cooling to the initial conditions we add a period of gradient flow evolution to starting at some timestep $\eta_{cool}$ (measured in conformal time) up until timestep $\eta=1.0$. Numerically, gradient flow evolution is obtained by taking the equations of motion and setting the accelerations to zero,
\begin{equation}
\dot{\phi} = D^jD_j\phi -\frac{\lambda}{2} (|\phi|^2 - 1) 
\end{equation}
\begin{equation}
\dot{F}_{0j} = \partial_j F_{ij} -2a^2 e^2 Im[\phi^* D_j \phi]\,.
\end{equation}

\begin{figure*}
\includegraphics[width=1.0\columnwidth]{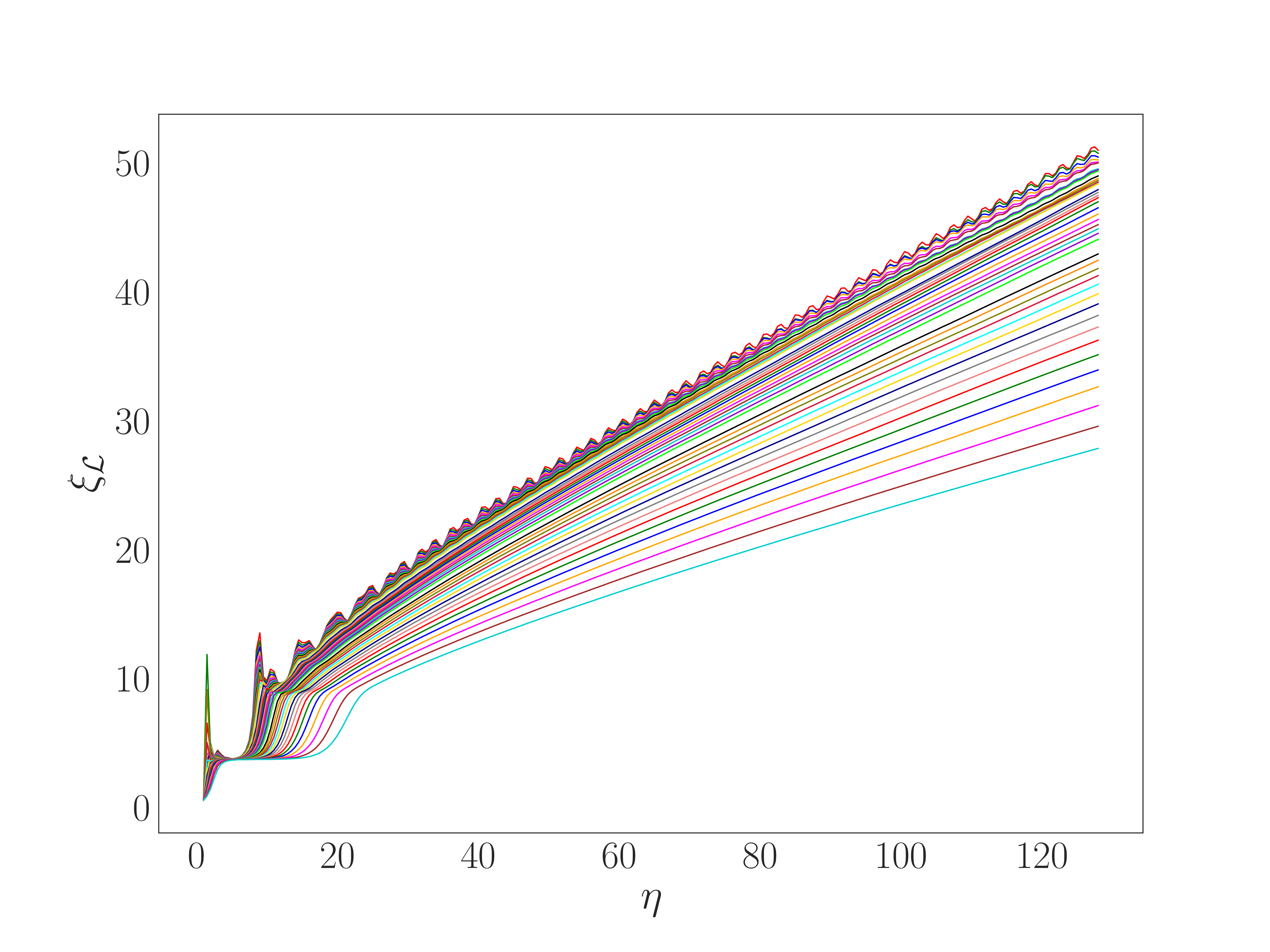}
\includegraphics[width=1.0\columnwidth]{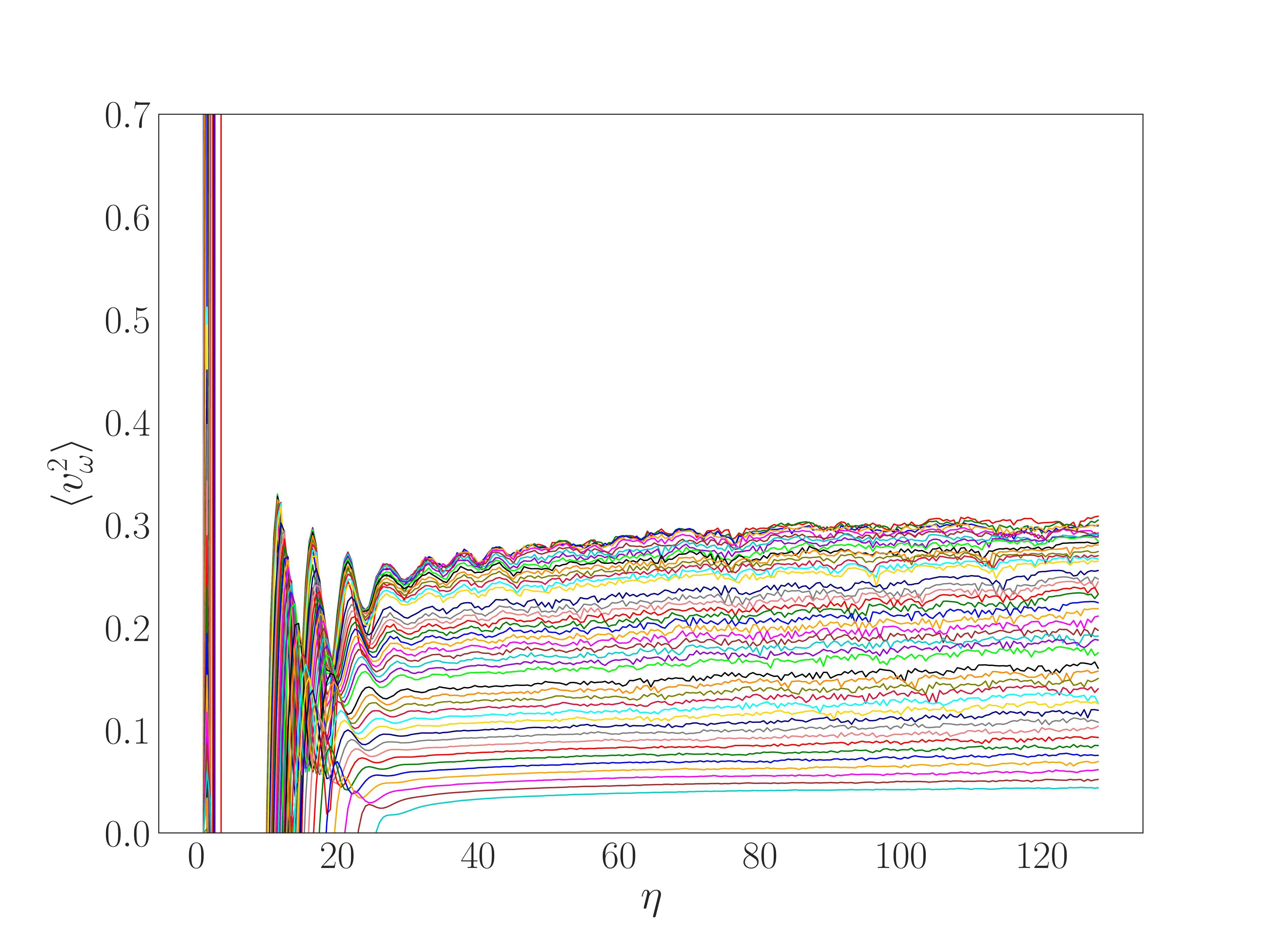}
\includegraphics[width=1.0\columnwidth]{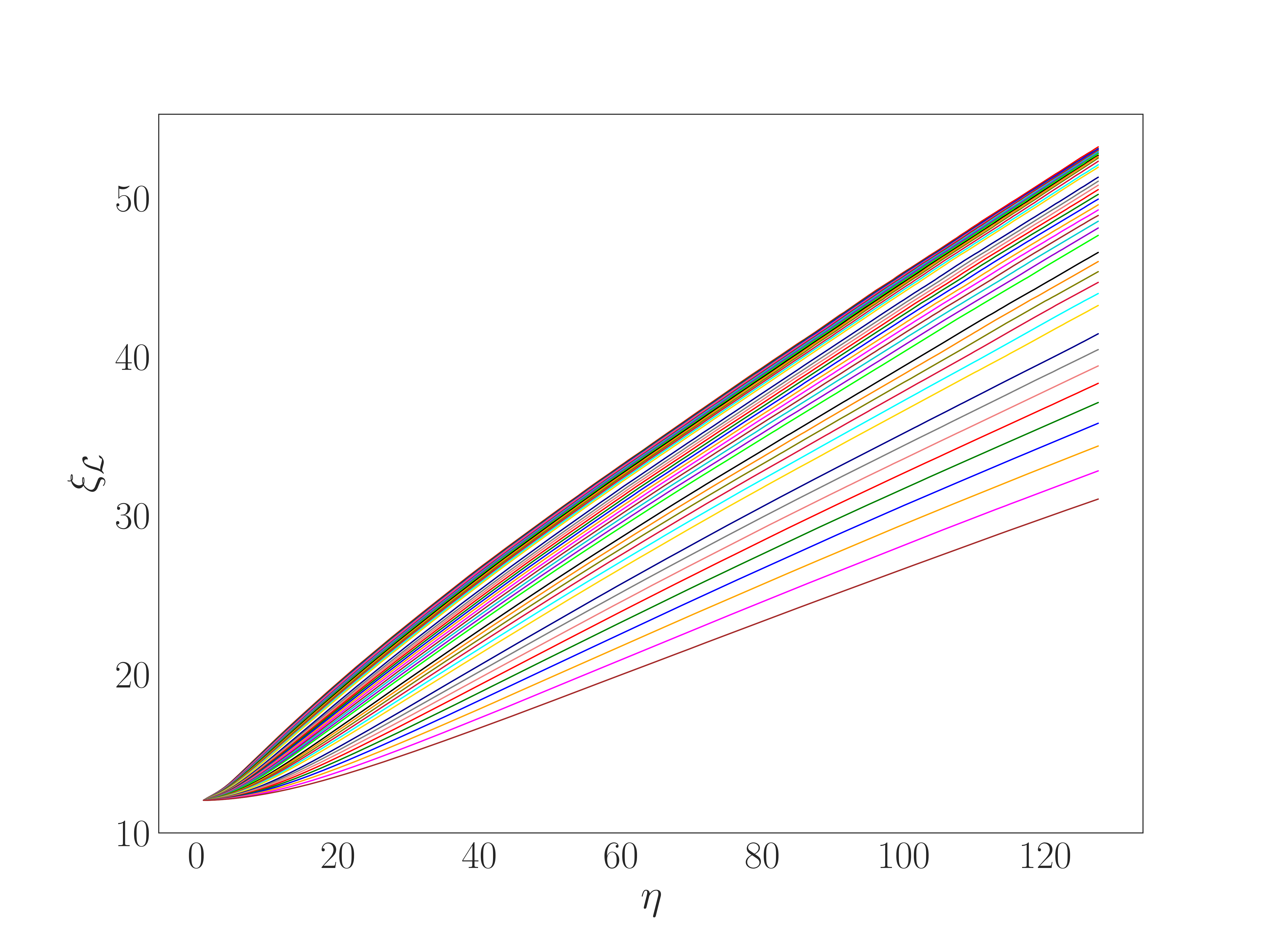}
\includegraphics[width=1.0\columnwidth]{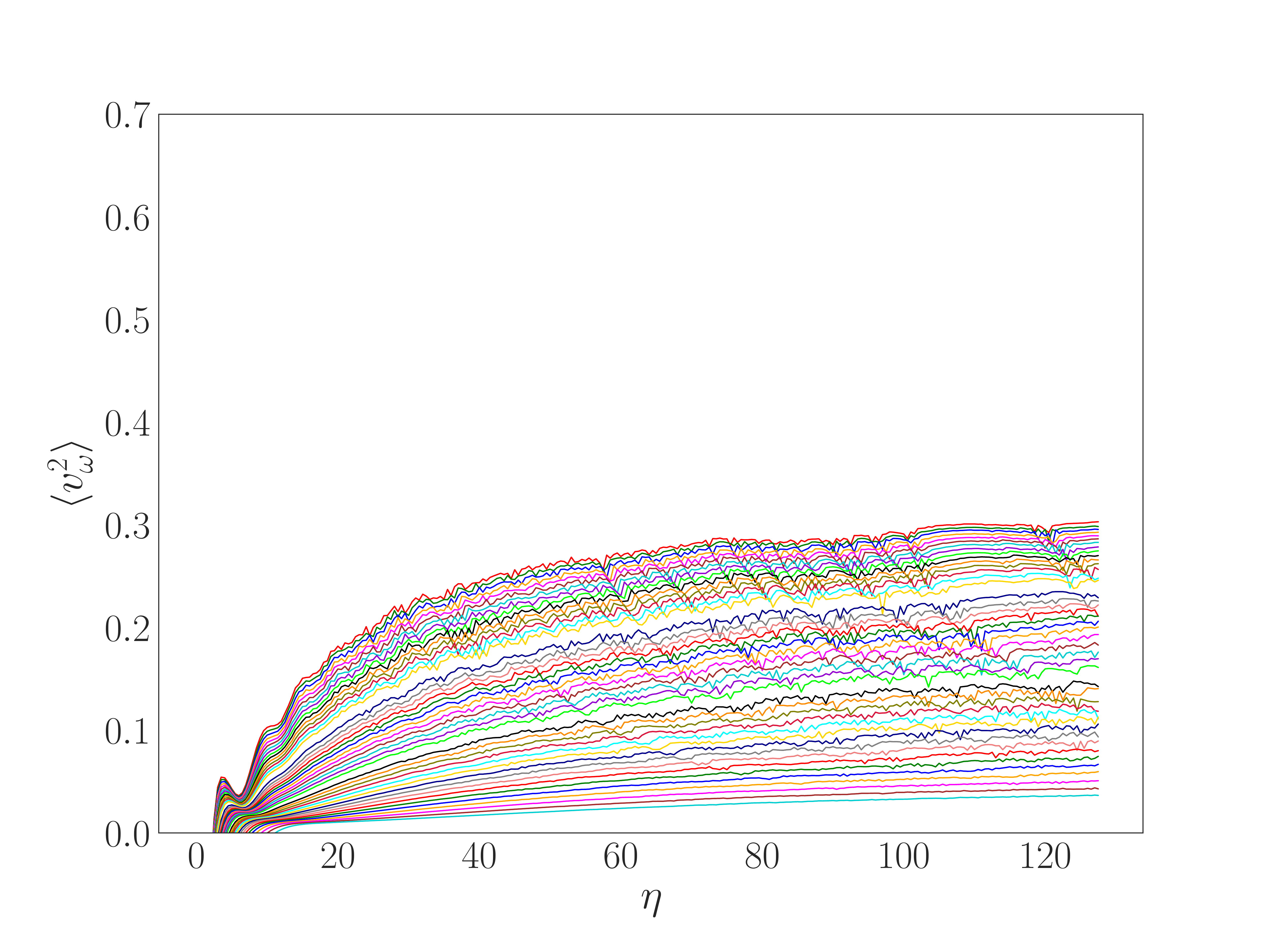}
\includegraphics[width=1.0\columnwidth]{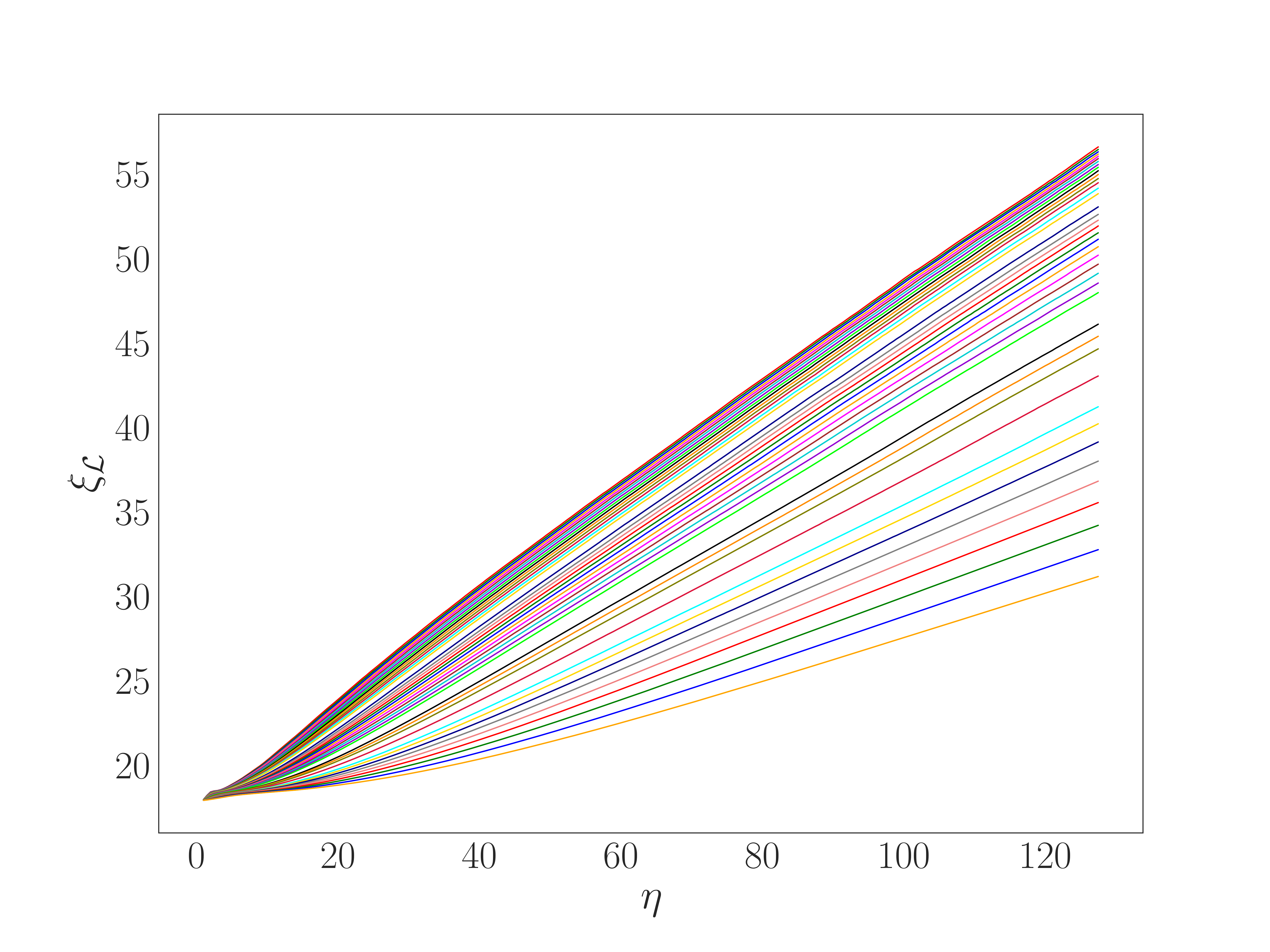}
\includegraphics[width=1.0\columnwidth]{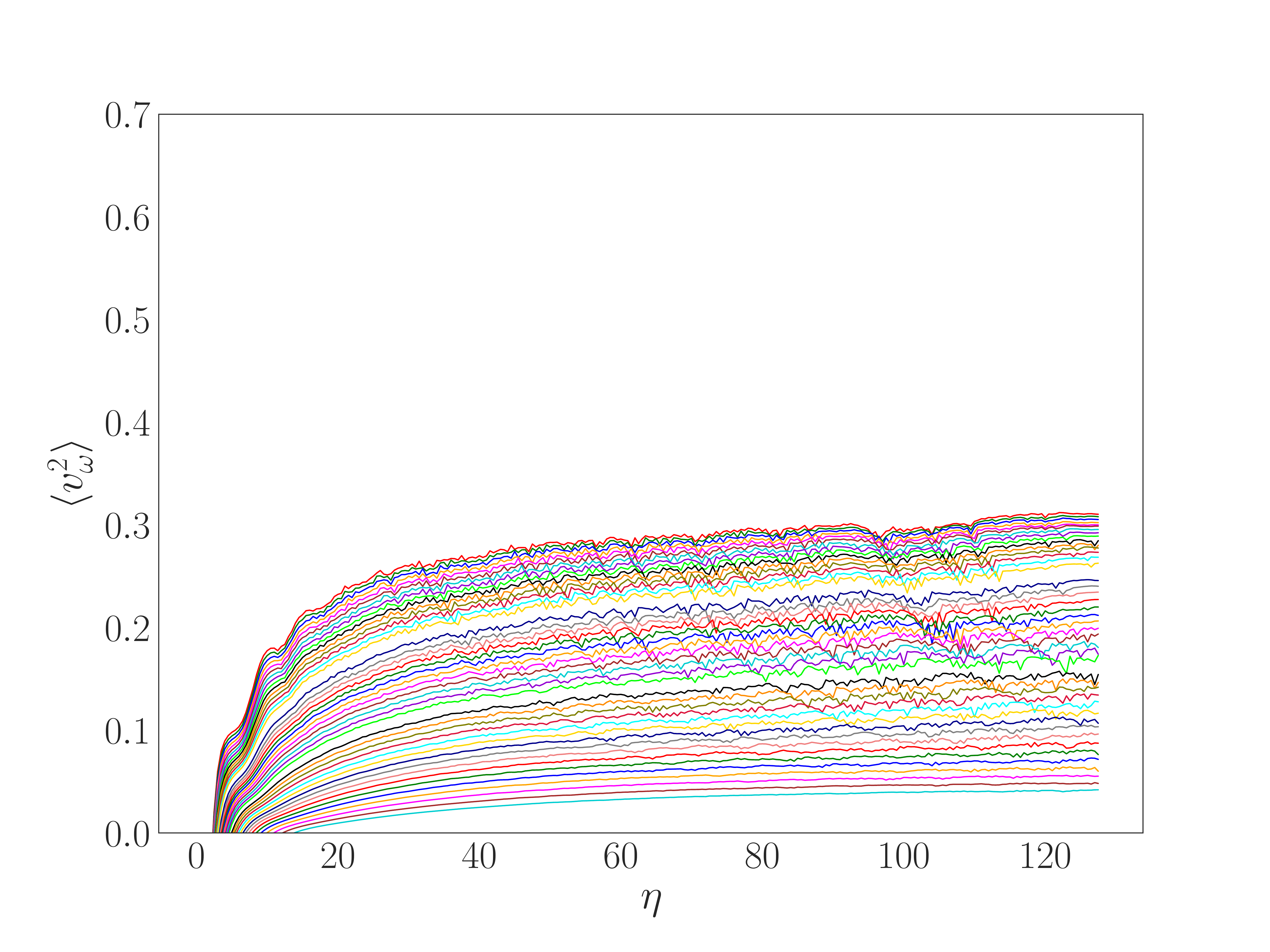}
\caption{The evolution of mean string separation according to the Lagrangian estimator (left panels) and the mean velocity squared according to equation of state estimator (right panels), averaged for sets of 12 runs at each expansion rate in the range $[0.50, 0.95]$. The top panel shows the results for the Hot case (standard case, without cooling), while the middle and bottom panels show the Warm and Cold cases. Low expansion rates are at the top of the panels while high expansion rates are at the bottom of the panels. All simulations have box sizes $512^3$ with constant comoving width (PRS algorithm).\label{fig1}}
\end{figure*}

The effect of such a period of cooling is to remove the large gradients present in the initial field configuration. This effect is confirmed by observing that the thermal oscillations visibly present, for low enough expansion rates, in the Lagrangian estimator, defined as
\begin{align}\label{diagxiLang}
  \xi_\mathcal{L} = \sqrt{ \frac{-\mu V}{\sum_x \mathcal{L}_x} }\,
\end{align}
(where $V$ is the box volume and $\mu$ the string tension) and in the equation of state velocity estimator defined above in Eq. \ref{diagvel}, also disappear, as shown in Fig. \ref{fig1}. Note that while in what follows we use the winding length estimator for the calibration, we display in Fig. \ref{fig1} the Lagrangian estimator precisely because this is the one for which the effects of the thermal oscillations are more obvious.

Given that we wish to observe the effects of varying degrees of cooling on the evolution of the network and the corresponding effects on the parameters of the VOS model, we will take the same 12 random initial conditions for each of our sets of simulations, which like in \cite{Correia:2019bdl} we do for 43 different expansion rates. Specifically we assume power law dependencies for the scale factor $a\propto t^m$, where $t$ is physical time and the constant values of $m$ are in the range $[0.50,0.95]$ (and are explicitly given in the next section). The radiation and matter era cases correspond to the choices $m=1/2$ and $m=2/3$, and are among the simulated expansion rates.

We simulate each of these sets under three different cooling scenarios:
\begin{itemize}
\item The standard case, without any artificial cooling applied, where the simulations start at our canonical choice of conformal time $\eta=1$; for simplicity, in what follows we refer this as the \textit{Hot} case. 
\item A small amount of cooling, with initial conditions chosen such that the gradient flow dissipation period starts at an effective $\eta_{cool}=-10.0$ and ends at $\eta=1$; in what follows we refer to this as the \textit{Warm} case.
\item A more significant amount of cooling, with initial conditions chosen such that the gradient flow dissipation period starts at an effective $\eta_{cool}=-50.0$ and ends at $\eta=1$; in what follows we refer to this as the \textit{Cold} case.
\end{itemize}

We emphasize that in all cases the cosmological evolution starts only at $\eta = 1.0$, and this evolution is exactly the same for all 43 expansion rates. The only difference is therefore in the initial condition boxes at $\eta = 1.0$, including the amount of radiation in each of them, which is expected to depend on the amount of cooling. It is for this period of cosmological evolution that we present the results in the following sections.

\section{\label{compar}Comparing calibrations}

Our Abelian-Higgs string simulations can be used to calibrate the VOS model \cite{Martins:1996jp}, an extension of which has recently been discussed in \cite{Correia:2019bdl}. Its evolution equations are
\begin{equation}
2\frac{dL}{dt} = 2HL(1+v^2) + F(v)
\end{equation}
\begin{equation}
\frac{dv}{dt} = \bigg( 1-v^2 \bigg) \bigg( \frac{k(v)}{L} -2Hv \bigg)
\end{equation}
where $L$ is the average correlation length (or equivalently within the one-scale approximation, the string separation) and $v$ the root-mean square velocity of a network of cosmic strings, $H$ is the Hubble parameter, $t$ is physical time, and $k(v)$ and $F(v)$ are two velocity-dependent parameters, respectively known as the momentum parameter (which encodes small-scale structure on the strings) and the energy loss term---where energy loss is understood to refer to physical mechanism other than Hubble damping. The momentum parameter, originally assumed to have the semi-analytic form \cite{Martins:2000cs}
\begin{equation}
    k(v) =  \frac{2\sqrt{2}}{\pi} \frac{1-8v^6}{1+8v^6} 
\end{equation}
has been extended to \cite{Rybak1}
\begin{equation}
    k(v) =  k_0\frac{1-(qv^2)^{\beta}}{1+(qv^2)^{\beta}}.
\end{equation}
where $k_0$, $q$ and $\beta$ are free parameters to be determined from numerical simulations. Clearly there will be some degeneracies between these parameters, which can only be broken by using considerable numbers of simulations, and specifically by having simulations with different expansion rates $m$, which will flesh out the velocity dependence of the various physical mechanisms impacting the evolution of the network.

Similarly, the energy loss function was extended to take into account two components,
\begin{equation}
    F(v) = cv + d[k_0-k(v)]^r
\end{equation}
where the linear term describes energy losses occurring through loop production while the second term is associated with losses by radiative processes (in other words, scalar and gauge radiation). Again, $d$, $r$ and $c$ are free-parameters. Thus this extended VOS model has a total of 6 free parameters, but as our previous work demonstrates \cite{Correia:2019bdl} (and will be further illustrated below) extant simulations can provide a calibration for all of them.

For our present purposes we convert the VOS to comoving coordinates, which are the ones used in the numerical simulations,
\begin{equation}
    \frac{d\xi}{d\eta} = \frac{m}{(1-m)\eta} v^2 +F(v)
    \end{equation}
    \begin{equation}
    \frac{dv}{d\eta} = (1-v^2) \bigg[ \frac{k(v)}{\xi}  - \frac{2m v}{(1-m)\eta} \bigg]
\end{equation}
We generically expect the following scaling laws
\begin{equation}
\xi\propto (\eta-\eta_0)^\mu
\end{equation}
\begin{equation}
v\propto \eta^\nu\,,
\end{equation}
with the scaling exponents having the specific values of $\mu=1$ and $\nu=0$ once the network has reached the scaling regime. The quality of scaling is also measured from the scaling exponents $\mu$ and $\nu$, with the fitting range chosen in order to ensure that they are as close as possible to the asymptotic scaling values, for as many expansion rates as possible (the limiting factor is the scaling of the velocities, which typically differ maximally by about ten percent). For the Hot case we rely on the data already reported in \cite{Correia:2019bdl} (where the fitting range $\eta \in [80, 128]$ was used), while for the Warm and Cold cases we report the result of new production runs with a more narrow fitting range $\eta \in [100,128]$. The fact that the network takes longer to reach scaling in the presence of cooling may seem counter-intuitive, but it comes from the requirement that both exponents ($\mu$ and $\nu$) are sufficiently close to the scaling values: while the mean string separation does reach scaling faster when cooling is applied (in agreement with previous reports in the literature) we find that this is not the case for the velocities. The relevant scaling exponents and network parameters of these simulations are listed in Table \ref{table1} and Table \ref{table2}, where for convenience we have defined
\begin{equation}
	\epsilon = \frac{\xi}{(1-m)(\eta- \eta_0)}\,.
\end{equation}

\begin{table}[h!]
\label{table1}
\begin{tabular}{| c | c c | c c |}
\hline
$m$ & $\mu$ & $\nu$ & $\epsilon$ & $\sqrt{\langle v^2 \rangle}$ \\
\hline
0.5 & 0.005$\pm$0.001 & 0.130$\pm$0.006 & 0.572$\pm$0.049 & 0.553$\pm$0.012\\ 
0.51 & 0.004$\pm$0.001 & 0.121$\pm$0.006 & 0.581$\pm$0.047 & 0.550$\pm$0.012\\ 
0.52 & 0.004$\pm$0.001 & 0.120$\pm$0.006 & 0.595$\pm$0.049 & 0.548$\pm$0.012\\ 
0.53 & 0.004$\pm$0.001 & 0.118$\pm$0.006 & 0.606$\pm$0.051 & 0.545$\pm$0.013\\ 
0.54 & 0.004$\pm$0.001 & 0.121$\pm$0.006 & 0.621$\pm$0.050 & 0.543$\pm$0.012\\ 
0.55 & 0.004$\pm$0.001 & 0.134$\pm$0.006 & 0.639$\pm$0.051 & 0.541$\pm$0.012\\ 
0.56 & 0.004$\pm$0.001 & 0.141$\pm$0.005 & 0.652$\pm$0.051 & 0.538$\pm$0.012\\ 
0.57 & 0.004$\pm$0.001 & 0.142$\pm$0.006 & 0.669$\pm$0.051 & 0.534$\pm$0.013\\ 
0.58 & 0.004$\pm$0.001 & 0.139$\pm$0.006 & 0.689$\pm$0.053 & 0.531$\pm$0.014\\ 
0.59 & 0.004$\pm$0.001 & 0.144$\pm$0.007 & 0.705$\pm$0.054 & 0.526$\pm$0.014\\ 
0.6 & 0.003$\pm$0.001 & 0.154$\pm$0.006 & 0.725$\pm$0.053 & 0.523$\pm$0.014\\ 
0.61 & 0.003$\pm$0.001 & 0.159$\pm$0.007 & 0.741$\pm$0.054 & 0.519$\pm$0.014\\ 
0.62 & 0.004$\pm$0.001 & 0.163$\pm$0.007 & 0.755$\pm$0.056 & 0.514$\pm$0.014\\ 
0.63 & 0.003$\pm$0.001 & 0.167$\pm$0.007 & 0.769$\pm$0.052 & 0.509$\pm$0.015\\ 
0.64 & 0.004$\pm$0.001 & 0.157$\pm$0.008 & 0.789$\pm$0.058 & 0.503$\pm$0.015\\ 
0.6(6)& 0.004$\pm$0.001 & 0.165$\pm$0.009 & 0.834$\pm$0.064 & 0.487$\pm$0.016\\ 
0.68 & 0.004$\pm$0.001 & 0.182$\pm$0.009 & 0.870$\pm$0.062 & 0.480$\pm$0.016\\ 
0.69 & 0.004$\pm$0.001 & 0.182$\pm$0.010 & 0.895$\pm$0.064 & 0.473$\pm$0.016\\ 
0.7 & 0.004$\pm$0.001 & 0.166$\pm$0.011 & 0.918$\pm$0.065 & 0.466$\pm$0.016\\ 
0.71 & 0.004$\pm$0.001 & 0.134$\pm$0.013 & 0.942$\pm$0.066 & 0.459$\pm$0.016\\ 
0.72 & 0.004$\pm$0.001 & 0.105$\pm$0.014 & 0.974$\pm$0.071 & 0.452$\pm$0.017\\ 
0.73 & 0.004$\pm$0.001 & 0.092$\pm$0.014 & 1.003$\pm$0.073 & 0.445$\pm$0.016\\ 
0.74 & 0.004$\pm$0.001 & 0.073$\pm$0.014 & 1.037$\pm$0.075 & 0.438$\pm$0.016\\ 
0.75 & 0.004$\pm$0.001 & 0.074$\pm$0.015 & 1.078$\pm$0.083 & 0.431$\pm$0.017\\ 
0.76 & 0.004$\pm$0.001 & 0.053$\pm$0.014 & 1.119$\pm$0.082 & 0.424$\pm$0.016\\ 
0.77 & 0.004$\pm$0.001 & 0.036$\pm$0.016 & 1.160$\pm$0.084 & 0.417$\pm$0.015\\ 
0.78 & 0.004$\pm$0.001 & 0.020$\pm$0.017 & 1.203$\pm$0.085 & 0.408$\pm$0.016\\ 
0.8 & 0.004$\pm$0.001 & 0.038$\pm$0.018 & 1.295$\pm$0.093 & 0.390$\pm$0.017\\ 
0.82 & 0.004$\pm$0.001 & 0.067$\pm$0.019 & 1.407$\pm$0.100 & 0.371$\pm$0.016\\ 
0.83 & 0.003$\pm$0.001 & 0.087$\pm$0.018 & 1.469$\pm$0.098 & 0.361$\pm$0.015\\ 
0.84 & 0.003$\pm$0.001 & 0.092$\pm$0.019 & 1.534$\pm$0.104 & 0.351$\pm$0.014\\ 
0.85 & 0.003$\pm$0.001 & 0.122$\pm$0.017 & 1.598$\pm$0.110 & 0.340$\pm$0.013\\ 
0.86 & 0.003$\pm$0.001 & 0.140$\pm$0.016 & 1.663$\pm$0.117 & 0.328$\pm$0.013\\ 
0.87 & 0.003$\pm$0.001 & 0.158$\pm$0.014 & 1.736$\pm$0.125 & 0.317$\pm$0.012\\ 
0.88 & 0.003$\pm$0.001 & 0.161$\pm$0.016 & 1.824$\pm$0.127 & 0.303$\pm$0.011\\ 
0.89 & 0.003$\pm$0.001 & 0.147$\pm$0.016 & 1.928$\pm$0.127 & 0.291$\pm$0.011\\ 
0.9 & 0.003$\pm$0.001 & 0.145$\pm$0.015 & 2.043$\pm$0.128 & 0.278$\pm$0.01\\ 
0.91 & 0.003$\pm$0.001 & 0.127$\pm$0.013 & 2.176$\pm$0.152 & 0.263$\pm$0.009\\ 
0.92 & 0.004$\pm$0.001 & 0.095$\pm$0.011 & 2.339$\pm$0.181 & 0.249$\pm$0.008\\ 
0.93 & 0.004$\pm$0.001 & 0.093$\pm$0.011 & 2.544$\pm$0.204 & 0.234$\pm$0.007\\ 
0.94 & 0.003$\pm$0.001 & 0.096$\pm$0.009 & 2.820$\pm$0.199 & 0.218$\pm$0.005\\ 
0.95 & 0.002$\pm$0.001 & 0.081$\pm$0.009 & 3.190$\pm$0.183 & 0.202$\pm$0.004\\ 
\hline
\end{tabular}
\caption{Scaling exponents $\mu$ and $\nu$ and network parameters used for VOS calibration for the Warm initial conditions case. One-sigma statistical uncertainties, from averaging sets of 12 simulations, are reported throughout.}
\end{table}

\begin{table}[h!]
\label{table2}
\begin{tabular}{| c | c c | c c |}
\hline
$m$ & $\mu$ & $\nu$ & $\epsilon$ & $\sqrt{\langle v^2 \rangle}$ \\
\hline
0.5 & 0.005$\pm$0.001 & 0.085$\pm$0.003 & 0.560$\pm$0.047 & 0.542$\pm$0.012\\ 
0.51 & 0.005$\pm$0.001 & 0.082$\pm$0.004 & 0.572$\pm$0.047 & 0.539$\pm$0.012\\ 
0.52 & 0.005$\pm$0.001 & 0.086$\pm$0.004 & 0.585$\pm$0.049 & 0.536$\pm$0.013\\ 
0.53 & 0.005$\pm$0.001 & 0.090$\pm$0.004 & 0.597$\pm$0.051 & 0.532$\pm$0.013\\ 
0.54 & 0.005$\pm$0.001 & 0.093$\pm$0.004 & 0.610$\pm$0.053 & 0.529$\pm$0.013\\ 
0.55 & 0.005$\pm$0.001 & 0.096$\pm$0.004 & 0.626$\pm$0.054 & 0.526$\pm$0.014\\ 
0.56 & 0.005$\pm$0.001 & 0.101$\pm$0.004 & 0.640$\pm$0.055 & 0.522$\pm$0.015\\ 
0.57 & 0.005$\pm$0.001 & 0.099$\pm$0.004 & 0.657$\pm$0.057 & 0.518$\pm$0.015\\ 
0.58 & 0.005$\pm$0.001 & 0.103$\pm$0.004 & 0.673$\pm$0.057 & 0.514$\pm$0.015\\ 
0.59 & 0.005$\pm$0.001 & 0.104$\pm$0.004 & 0.688$\pm$0.058 & 0.510$\pm$0.015\\ 
0.6 & 0.005$\pm$0.001 & 0.106$\pm$0.004 & 0.705$\pm$0.061 & 0.506$\pm$0.016\\ 
0.61 & 0.005$\pm$0.001 & 0.105$\pm$0.005 & 0.723$\pm$0.061 & 0.502$\pm$0.016\\ 
0.62 & 0.005$\pm$0.001 & 0.107$\pm$0.005 & 0.741$\pm$0.063 & 0.497$\pm$0.016\\ 
0.63 & 0.005$\pm$0.001 & 0.108$\pm$0.005 & 0.757$\pm$0.062 & 0.492$\pm$0.016\\ 
0.64 & 0.005$\pm$0.001 & 0.110$\pm$0.005 & 0.774$\pm$0.064 & 0.486$\pm$0.016\\ 
0.6(6) & 0.005$\pm$0.001 & 0.107$\pm$0.005 & 0.833$\pm$0.066 & 0.472$\pm$0.016\\ 
0.68 & 0.004$\pm$0.001 & 0.105$\pm$0.005 & 0.864$\pm$0.068 & 0.464$\pm$0.016\\ 
0.69 & 0.004$\pm$0.001 & 0.112$\pm$0.005 & 0.889$\pm$0.070 & 0.458$\pm$0.016\\ 
0.7 & 0.004$\pm$0.001 & 0.110$\pm$0.005 & 0.916$\pm$0.071 & 0.452$\pm$0.017\\ 
0.71 & 0.004$\pm$0.001 & 0.108$\pm$0.006 & 0.943$\pm$0.071 & 0.445$\pm$0.016\\ 
0.72 & 0.004$\pm$0.001 & 0.113$\pm$0.005 & 0.973$\pm$0.075 & 0.438$\pm$0.017\\ 
0.73 & 0.004$\pm$0.001 & 0.114$\pm$0.005 & 1.005$\pm$0.075 & 0.431$\pm$0.017\\ 
0.74 & 0.004$\pm$0.001 & 0.107$\pm$0.006 & 1.035$\pm$0.076 & 0.423$\pm$0.017\\ 
0.75 & 0.004$\pm$0.001 & 0.108$\pm$0.006 & 1.067$\pm$0.077 & 0.415$\pm$0.017\\ 
0.76 & 0.004$\pm$0.001 & 0.108$\pm$0.006 & 1.101$\pm$0.078 & 0.407$\pm$0.018\\ 
0.77 & 0.004$\pm$0.001 & 0.111$\pm$0.006 & 1.137$\pm$0.082 & 0.398$\pm$0.017\\ 
0.78 & 0.004$\pm$0.001 & 0.112$\pm$0.006 & 1.177$\pm$0.084 & 0.390$\pm$0.016\\ 
0.8 & 0.003$\pm$0.001 & 0.122$\pm$0.006 & 1.266$\pm$0.088 & 0.372$\pm$0.016\\ 
0.82 & 0.003$\pm$0.001 & 0.152$\pm$0.007 & 1.362$\pm$0.092 & 0.353$\pm$0.015\\ 
0.83 & 0.003$\pm$0.001 & 0.168$\pm$0.008 & 1.414$\pm$0.096 & 0.343$\pm$0.015\\ 
0.84 & 0.003$\pm$0.001 & 0.197$\pm$0.008 & 1.470$\pm$0.100 & 0.332$\pm$0.014\\ 
0.85 & 0.003$\pm$0.001 & 0.212$\pm$0.008 & 1.529$\pm$0.105 & 0.321$\pm$0.014\\ 
0.86 & 0.003$\pm$0.001 & 0.216$\pm$0.008 & 1.591$\pm$0.110 & 0.309$\pm$0.013\\ 
0.87 & 0.003$\pm$0.001 & 0.222$\pm$0.007 & 1.656$\pm$0.115 & 0.298$\pm$0.012\\ 
0.88 & 0.003$\pm$0.001 & 0.228$\pm$0.007 & 1.729$\pm$0.120 & 0.285$\pm$0.011\\ 
0.89 & 0.003$\pm$0.001 & 0.234$\pm$0.006 & 1.812$\pm$0.127 & 0.272$\pm$0.010\\ 
0.9 & 0.003$\pm$0.001 & 0.228$\pm$0.006 & 1.908$\pm$0.136 & 0.259$\pm$0.009\\ 
0.91 & 0.003$\pm$0.001 & 0.227$\pm$0.005 & 2.016$\pm$0.154 & 0.245$\pm$0.008\\ 
0.92 & 0.003$\pm$0.001 & 0.228$\pm$0.004 & 2.147$\pm$0.166 & 0.230$\pm$0.007\\ 
0.93 & 0.003$\pm$0.001 & 0.221$\pm$0.003 & 2.311$\pm$0.177 & 0.215$\pm$0.006\\ 
0.94 & 0.003$\pm$0.001 & 0.227$\pm$0.003 & 2.517$\pm$0.185 & 0.200$\pm$0.005\\ 
0.95 & 0.002$\pm$0.001 & 0.240$\pm$0.003 & 2.774$\pm$0.184 & 0.183$\pm$0.004\\ 
\hline
\end{tabular}
\caption{Scaling exponents $\mu$ and $\nu$ and network parameters used for VOS calibration for the Cold initial conditions case. One-sigma statistical uncertainties, from averaging sets of 12 simulations, are reported throughout.}
\end{table}

\begin{figure*}
\includegraphics[width=1.0\columnwidth]{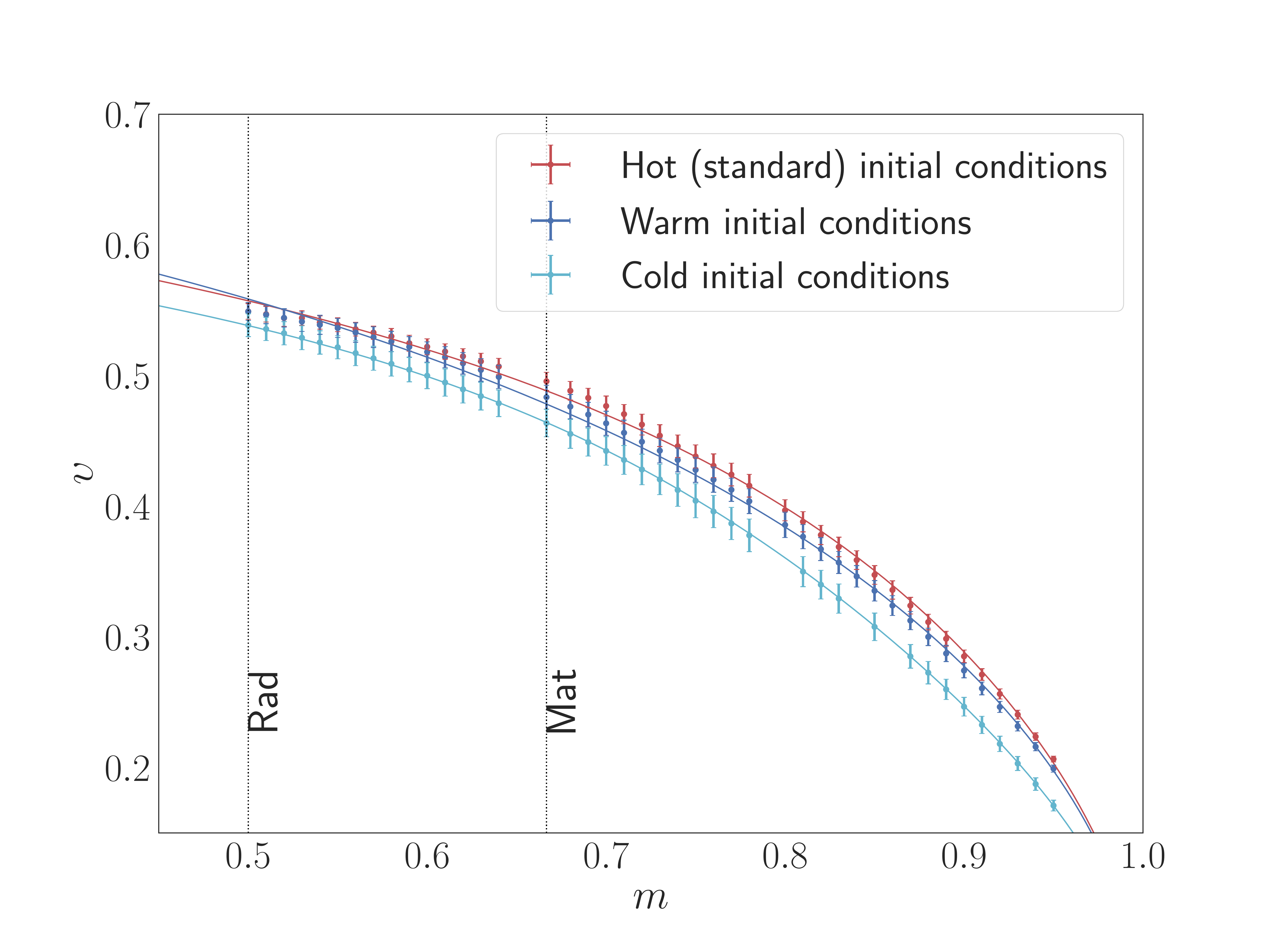}
\includegraphics[width=1.0\columnwidth]{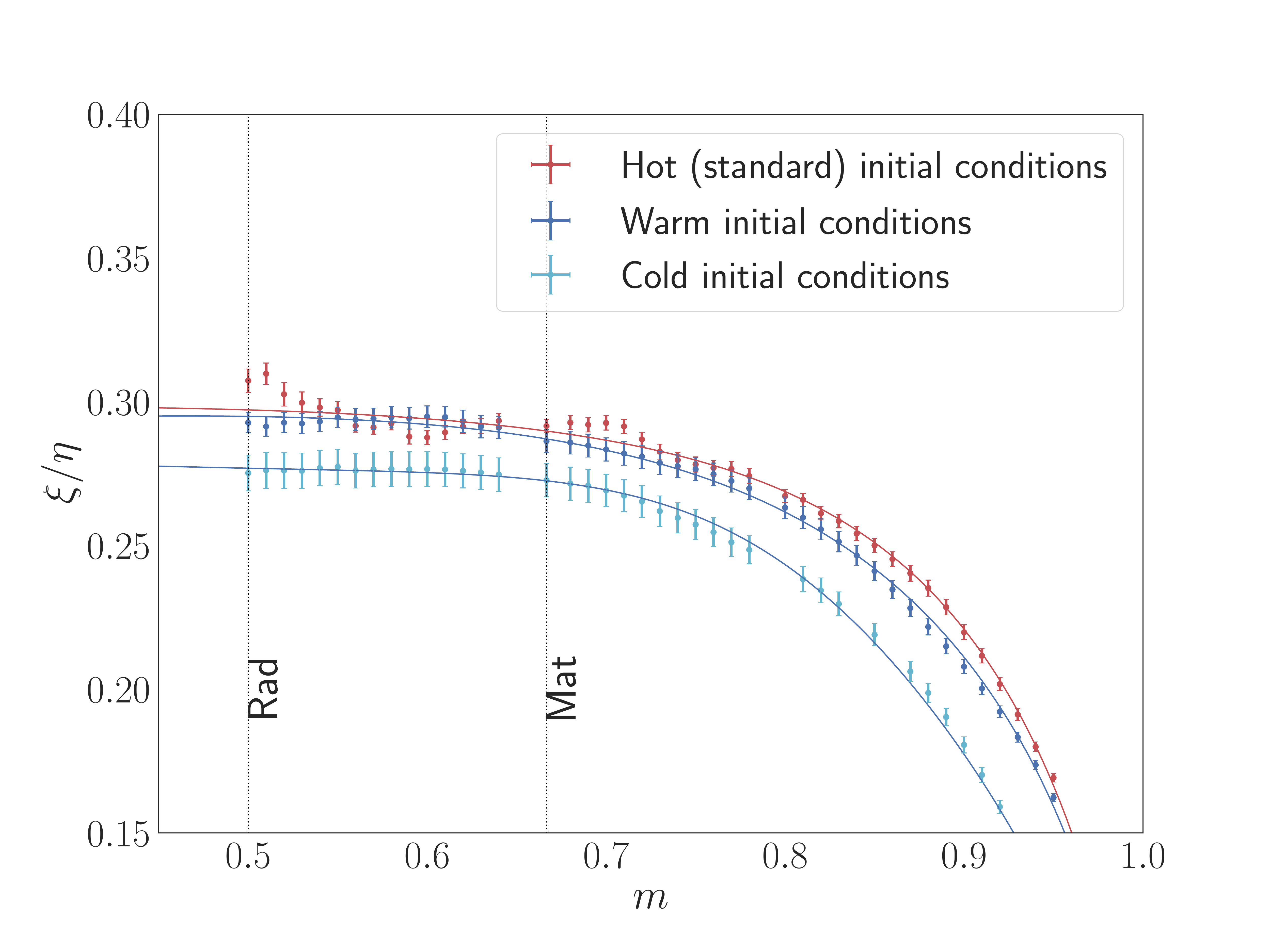}
\includegraphics[width=1.0\columnwidth]{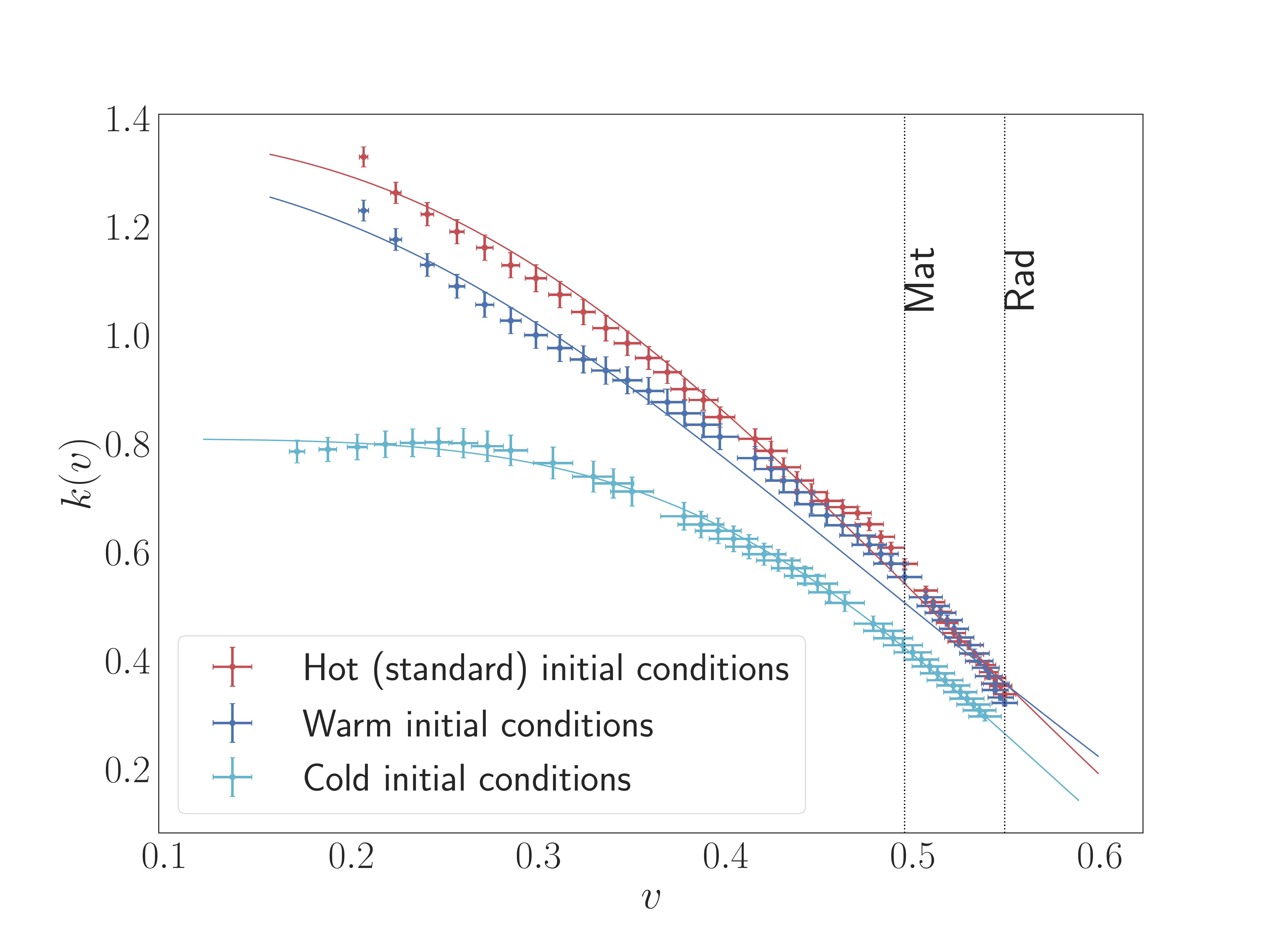}
\includegraphics[width=1.0\columnwidth]{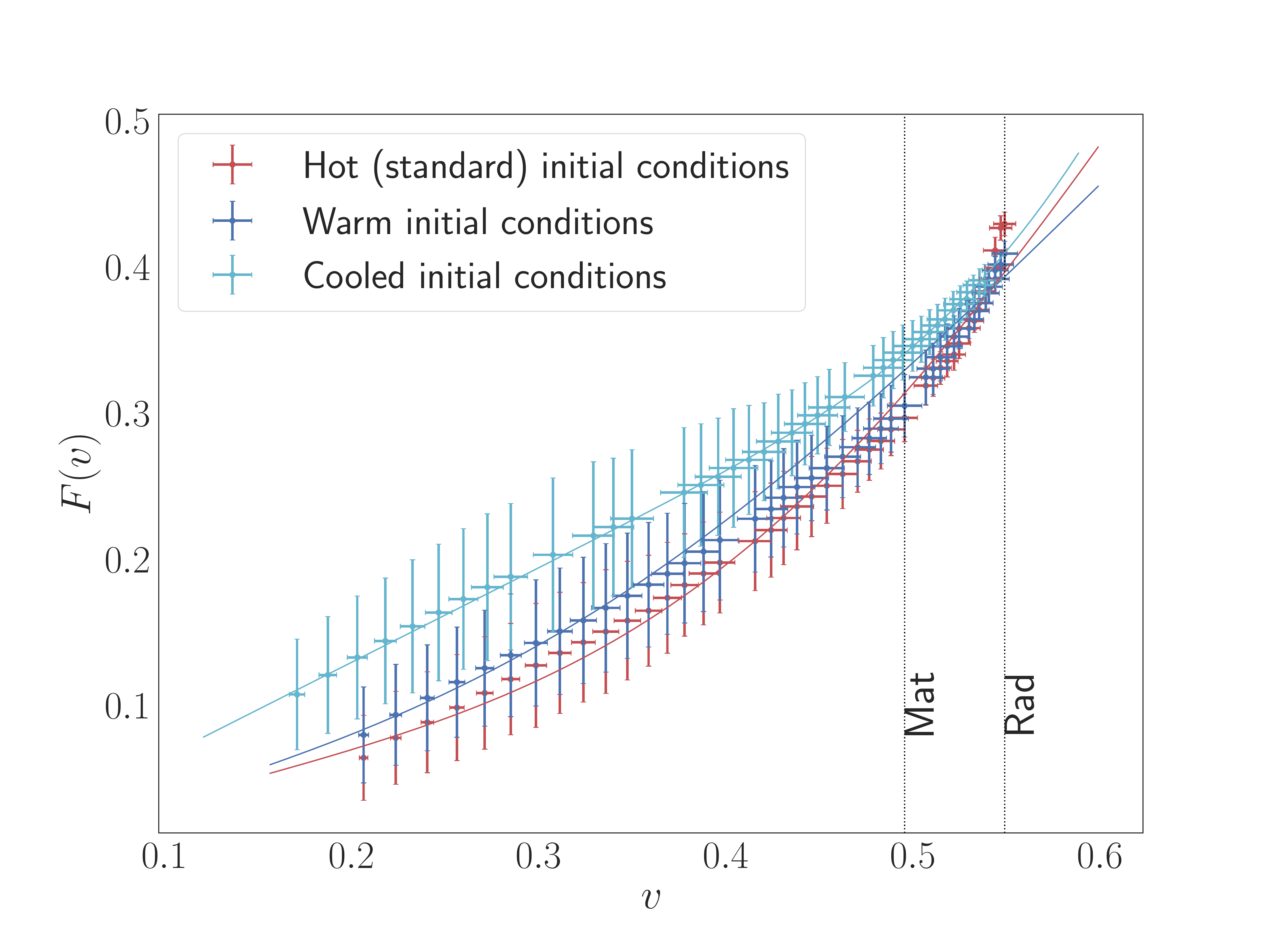}
\caption{Top panels: the string network average velocity and dimensionless comoving string separation, $v=\sqrt{\langle v^2 \rangle}$ and $\xi/\eta$, respectively in the left and right panels, for the three cooling scenarios. Bottom panels: The momentum parameter and the energy loss function (left and right panels respectively) for the same cooling scenarios. In all cases the error bars are the statistical uncertainties from averaging over 12 simulations with different initial conditions, and the solid line is is the prediction from the VOS model, with the calibrated parameters listed in Table III. For convenience the values corresponding to simulations in the radiation and matter eras have been highlighted.\label{fig2}}
\end{figure*}

These allow us to obtain the calibrated parameters from the measured quantities from the simulations (which are summarized in the top panels of Fig. \ref{fig2}), using standard bootstrap methods as described in \cite{Correia:2019bdl,Rybak1}. The results of this analysis for our three cooling scenarios are summarized in Table III. Additionally, one can also invert the VOS equations to obtain expressions for the momentum parameter or the energy loss function,
\begin{equation}\label{eq:vos1}
    F(v) = 2 \epsilon [1-m (1-v_0^2)]
\end{equation}
\begin{equation}\label{eq:vos2}
    k(v) = 2m \epsilon v_0
\end{equation}
and directly obtain both functions from the simulation outputs; these are shown in the bottom panels of Fig. \ref{fig2}. Note that while re-writing the VOS model in the more compact form above we are interested in the slope of $\eta$ for model calibration purposes, and this is approximated traditionally by $\xi/\eta$. In our case, in the quantitative fitting to the simulation data we use the generalized definition $\xi/(\eta-\eta_0)$ which, given the expected scaling law of our simulations $\eta \propto (\eta - \eta_0)$ requires an initial conditions dependent offset $\eta_0$.

\begin{table*}
\label{table3}
\begin{tabular}{| c | c  c  c  c  c  c | c |}
\hline
Case & d & r & $\beta$ & $k_0$ & q & c & Reference\\
\hline
Hot & 0.21$\pm$0.01 & 1.85$\pm$0.11 & 1.46$\pm$0.07 & 1.37$\pm$0.07 & 2.30$\pm$0.04 & 0.34$\pm$0.02 & \citep{Correia:2019bdl} \\
Warm & 0.26$\pm$0.01 & 1.58$\pm$0.10 & 1.29$\pm$0.06 & 1.21$\pm$0.06 & 2.05$\pm$0.04 & 0.36$\pm$0.03 & This work \\
Cold & 0.17$\pm$0.01 & 1.64$\pm$0.09 & 1.91$\pm$0.03 & 0.97$\pm$0.03 & 2.38$\pm$0.02 & 0.56$\pm$0.01 & This work \\
\hline
\end{tabular}
\caption{Calibrated VOS model parameters for our three cooling scenarios: Hot (standard), Warm and Cold initial conditions. These were obtained through the previously used bootstrap methods.}
\end{table*}

The results are in agreement with expectations, considering that the main dynamical effect of the cooling is to remove the thermal oscillations. Indeed in the Warm case (where cooling is just enough to remove the oscillations present in the Lagrangian estimator), we can expect the calibration to remain mostly unchanged with possibly better agreement between model prediction and measured values. Overall we do see this, and the effects on the calibrated VOS model parameters are not statistically significant, considering that there are degeneracies in the model parameters and that the estimate of these parameter uncertainties may be somewhat optimistic (in other words, the error bars in Table III may be underestimated). We will revisit this issue in the following section.

On the other hand, in the Cold case the differences are clear and statistically significant. For example the amount of small-scale structure on the strings is reduced, which is manifest both by looking at animations of these simulations or, more quantitatively, in the fact that the $k_0$ parameter is reduced to below unity: indeed in the Hot, Warm and Cool cases the best-fit values of $k_0$ are respectively 1.37, 1.21 and 0.97. Note that this is a subtle effect that may be missed in lower resolution simulations, since the average velocities are not significantly affected, and even the effects of the string density (or characteristic length) are not dramatic.

It's also interesting to note that the fitted value of the loop chopping parameter $c$ clearly increases with the amount of cooling: from $c=0.34$ in the hot case to $c=0.56$ in the cold case. Again this is to be expected, if one recalls that the energy loss term contains two terms (with different velocity dependencies) which model loop production and radiation losses. We therefore interpret this as an indication that the analytic model is correctly identifying the reduced amount of radiation in the box (through its effects on the string velocities) and therefore prefers a larger loop production term. In other words, this suggests that at least qualitatively this modelling approach is appropriate.
 
\section{\label{calibs}An improved calibration pipeline}

The calibration of the VOS model, in its extended six-parameter version, is a non-trivial statistical task. For this reason, and in preparation for a larger set of $4096^3$ and $8192^3$ simulations that are currently in progress, we have also done robustness tests and implemented extensions of our model calibration pipeline. In this section we describe these in more detail. As a specific test of the new methodology, we discuss its impact one the conclusions of the analysis in the previous section, which has been done with our previous pipeline.

\subsection{Uncertainty propagation}

In our previous pipeline the VOS model parameters were obtained by bootstrap methods. This relied on the average values of the string separation (or density) and velocity, with the averages being over each set of 12 simulations, but this did not explicitly take into account the uncertainties in these averages (that is, the standard deviations). While these uncertainties are typically small, they are not independent of the expansion rate: specifically, for $\epsilon$ they grow with $m$, while for the velocities the behaviour is less uniform. It is therefore important to check whether these uncertainties impact the results, especially considering that one expects degeneracies between some of the parameters.

Thus the first of our pipeline improvements is the introduction of full uncertainty propagation for the values of the string separation and velocity. For each of these the average and standard deviation obtained from the sets of simulations are input into an array from the uncertainties Python package. Moreover, we also statistically compute the offset $\eta_0$ for each run and, at each expansion rate, the mean offset and the standard deviation are stored in a similar array. From this point onward, the uncertainties are propagated automatically via this package, which is both more convenient and less error-prone.

\begin{figure*}
\includegraphics[width=1.0\columnwidth]{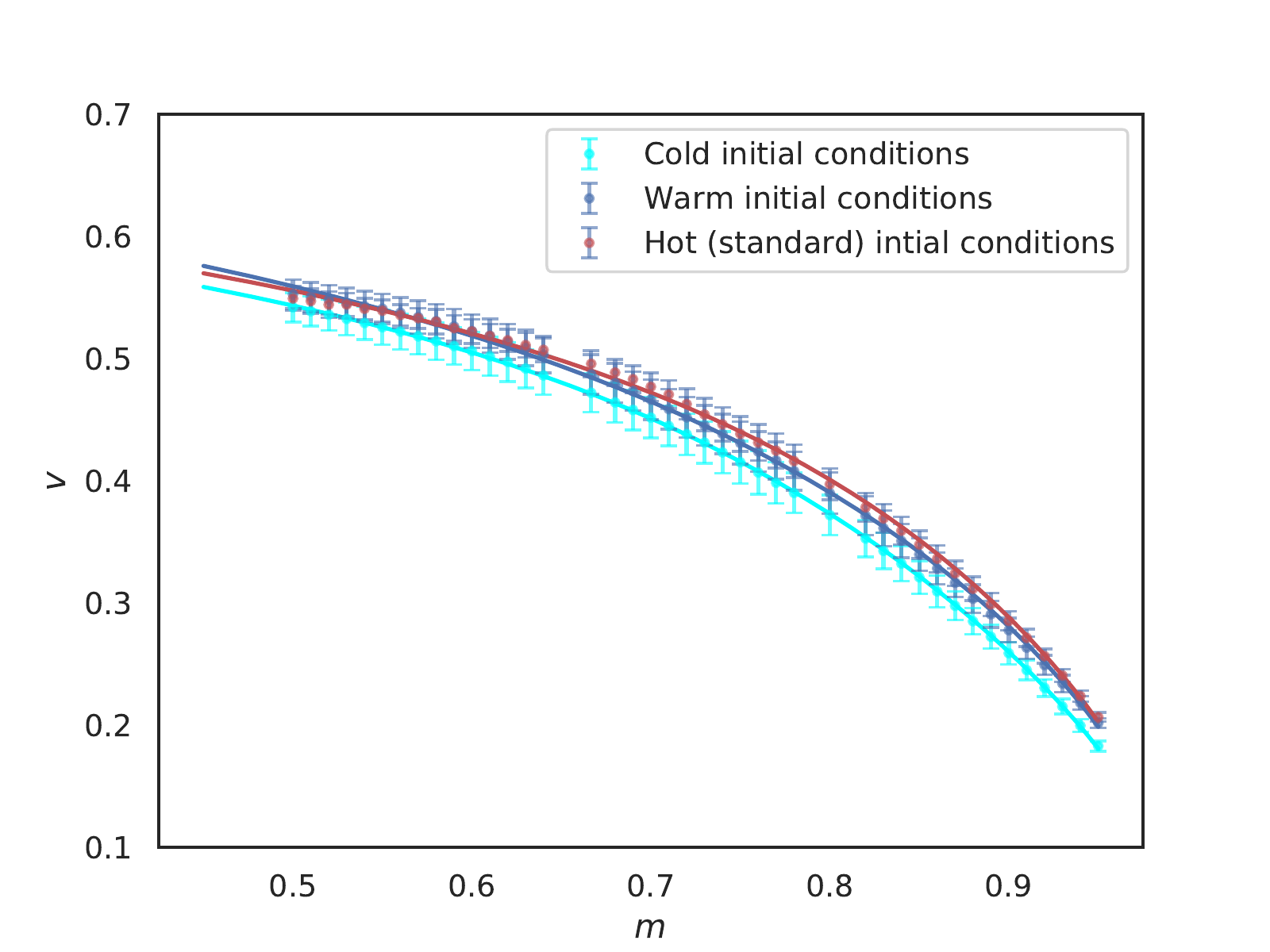}
\includegraphics[width=1.0\columnwidth]{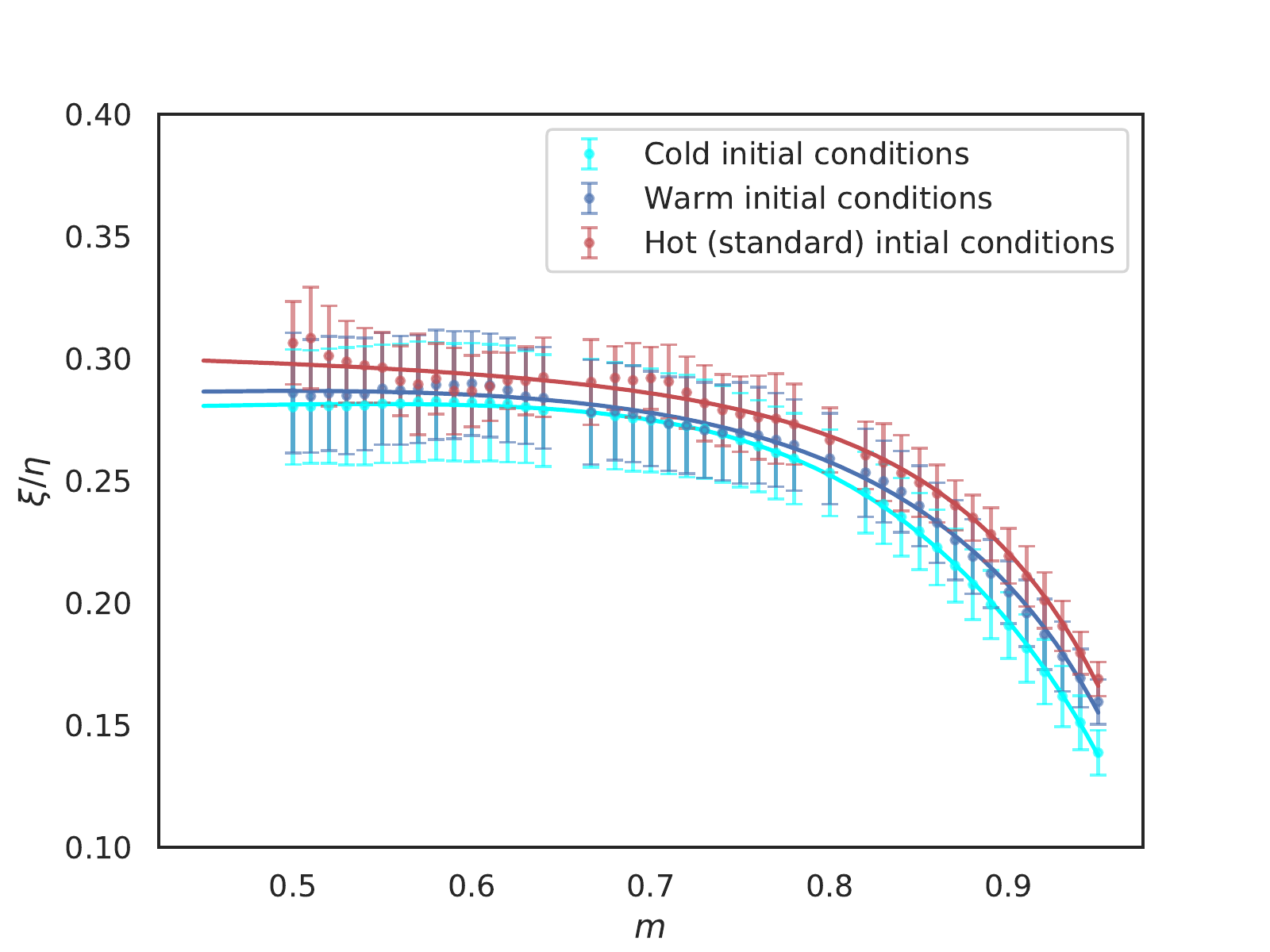}
\includegraphics[width=1.0\columnwidth]{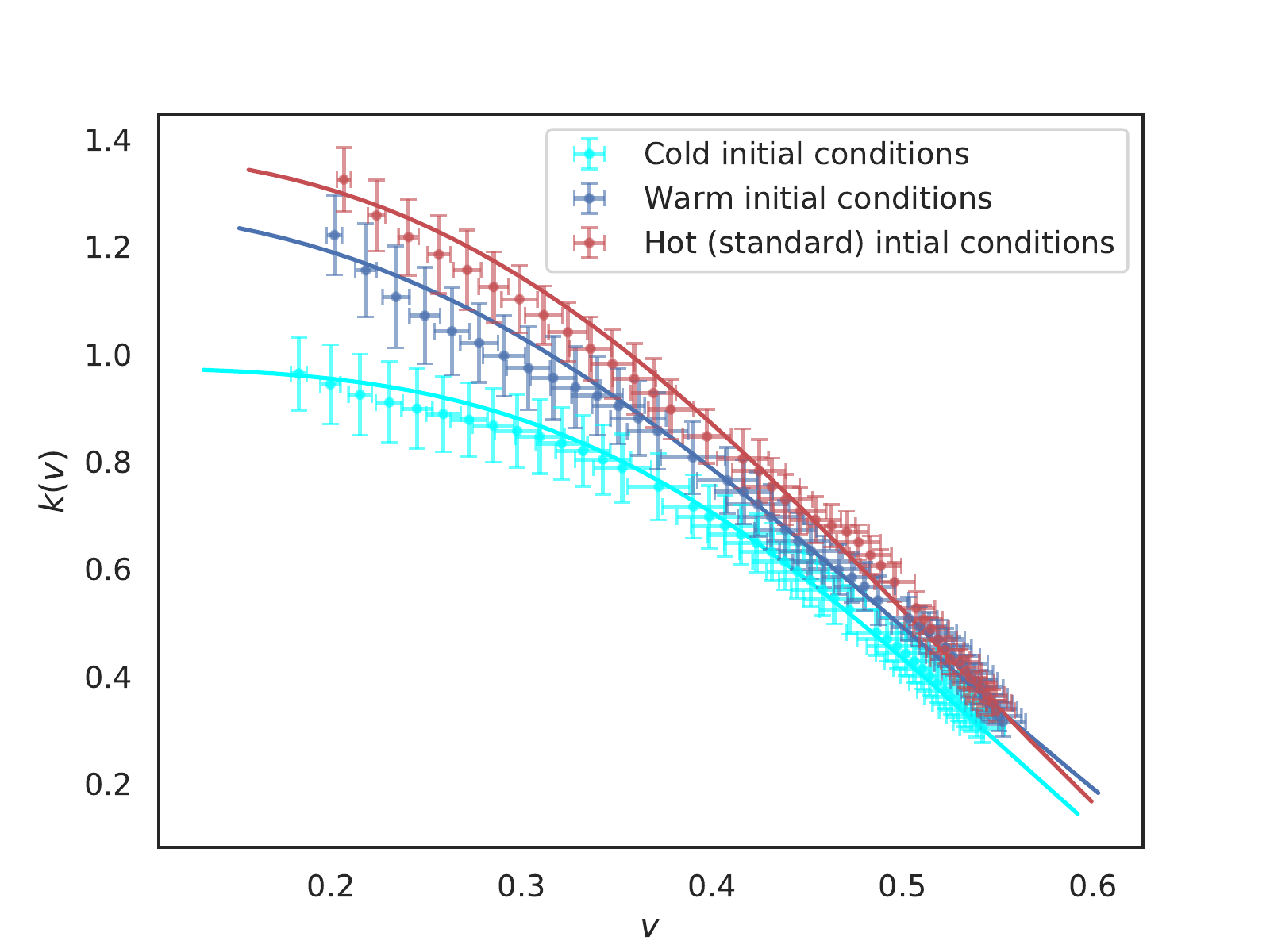}
\includegraphics[width=1.0\columnwidth]{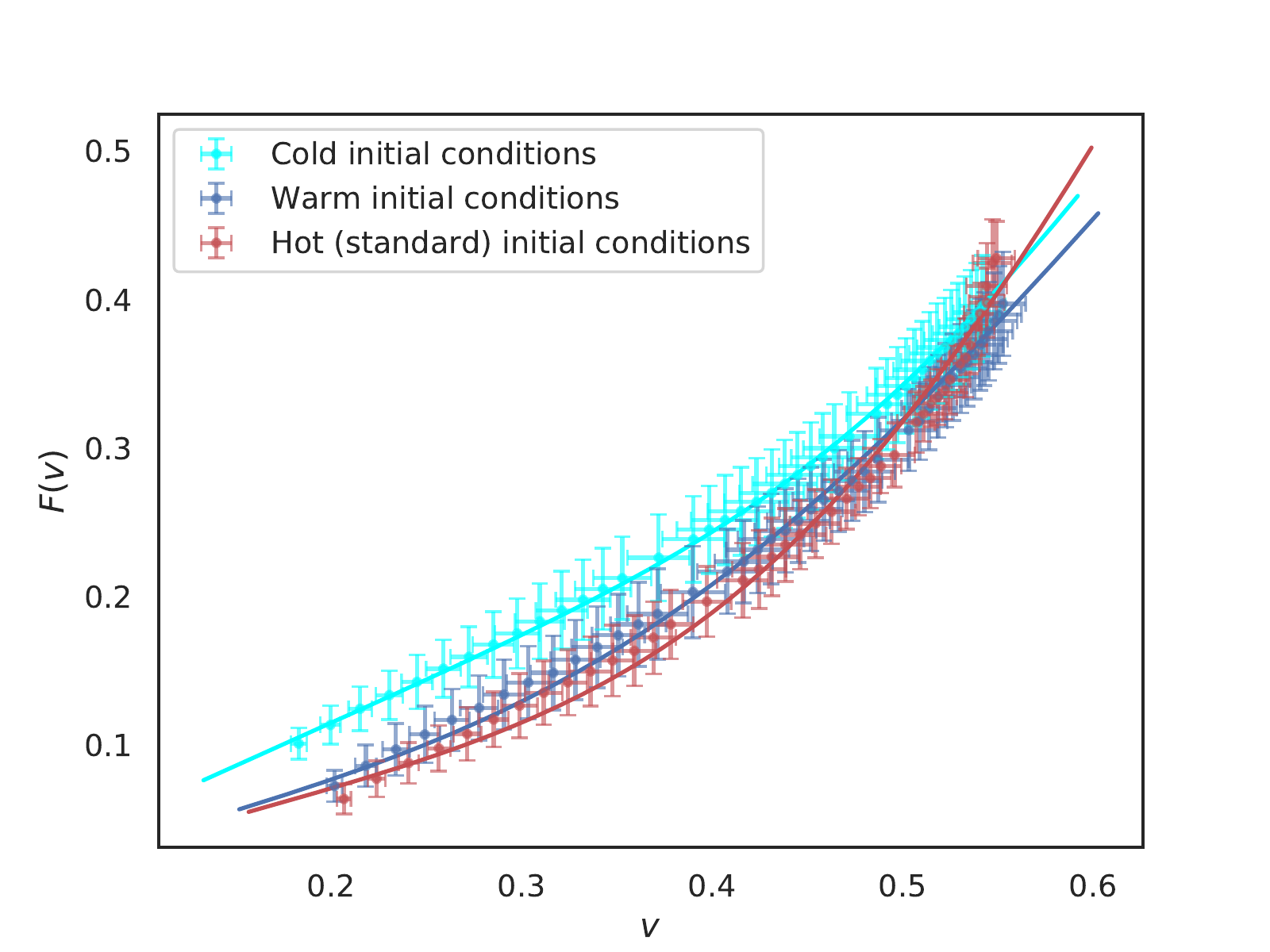}
\caption{Same as Figure \protect\ref{fig2}, but including the uncertainty propagation described in the main text and with the solid line now being the prediction from the VOS model with the calibrated parameters listed in Table IV.\label{fig3}}
\end{figure*}

We show the updated plots with this uncertainties in the top panels of Fig. \ref{fig3}. The calibrated parameters obtained with this analysis method are listed in Table IV, while the momentum parameter and the energy loss function are shown in the bottom panels of Fig. \ref{fig3}. Broadly speaking the uncertainties in $\xi / (\eta - \eta_0)$ increase and the uncertainties of $F(v)$ become larger at smaller expansion rates, while they are reduced at high expansion rates.

\begin{table*}
\label{table4}
\begin{tabular}{| c | c  c  c  c  c  c | c |}
\hline
Case & d & r & $\beta$ & $k_0$ & q & c & Reference\\
\hline
Hot & $0.20\pm0.01$ & $2.06\pm0.13$ & $1.54\pm0.06$ & $1.38\pm0.02$  & $2.38\pm0.03$ & $0.35\pm0.01$ & This work \\
Warm & $0.21\pm0.01$ & $1.68\pm0.12$ & $1.41\pm0.05$ & $1.27\pm0.02$ & $2.24\pm0.03$  & $0.37\pm0.01$ & This work \\
Cold & $0.19\pm0.01$ & $2.00\pm0.10$ & $1.95\pm0.03$ & $0.98\pm0.01$ & $2.45\pm0.01$  & $0.58\pm0.01$ & This work \\
\hline
\end{tabular}
\caption{Same as Table III, but including the uncertainty propagation described in the text.}
\end{table*}

Overall the conclusions taken in the previous section remain largely unchanged, especially for the more directly relevant parameters $c$, $d$ and $k_0$. Comparing Table IV with Table III we see that the changes are relatively small, and mostly within one or two standard deviations, although they tend to be slightly larger in the Cold case than in the Hot case. That said, the uncertainties in some of the model parameters actually decrease. While this can happen in our case due to correlations between several model parameters, this also suggests that a more robust calibration procedure and model parameter uncertainty estimation is desirable. We address this in the following sub-section.

\subsection{Bayesian Inference}

In order to improve the VOS calibration, and in particular the estimation of the uncertainties in the model parameters, we have implemented MCMC capabilities in our VOS calibration tool, specifically using emcee\footnote{\url{https://emcee.readthedocs.io/en/stable/}} \cite{emcee}. In addition to being a more robust estimation method, it has several advantages as a comparison point for our uncertainties previously obtained by the bootstraping minimization method---not only providing a check of our reasonable they are but also testing whether the minimum (best-fit) solution found via bootstraping is indeed a global mininum and identifying the limiting parameter dependencies. 

For our case we assume logarithmic probability density functions from uniform distributions for all priors. We use the $\chi^2$ statistic in order to compute the logarithm of the likelihood. We use 32 walkers and to be on the safe side, $10000$ steps. This proves sufficient to achieve convergence and a mean acceptance rate of around $0.4$.

\begin{figure*}
\includegraphics[width=\textwidth]{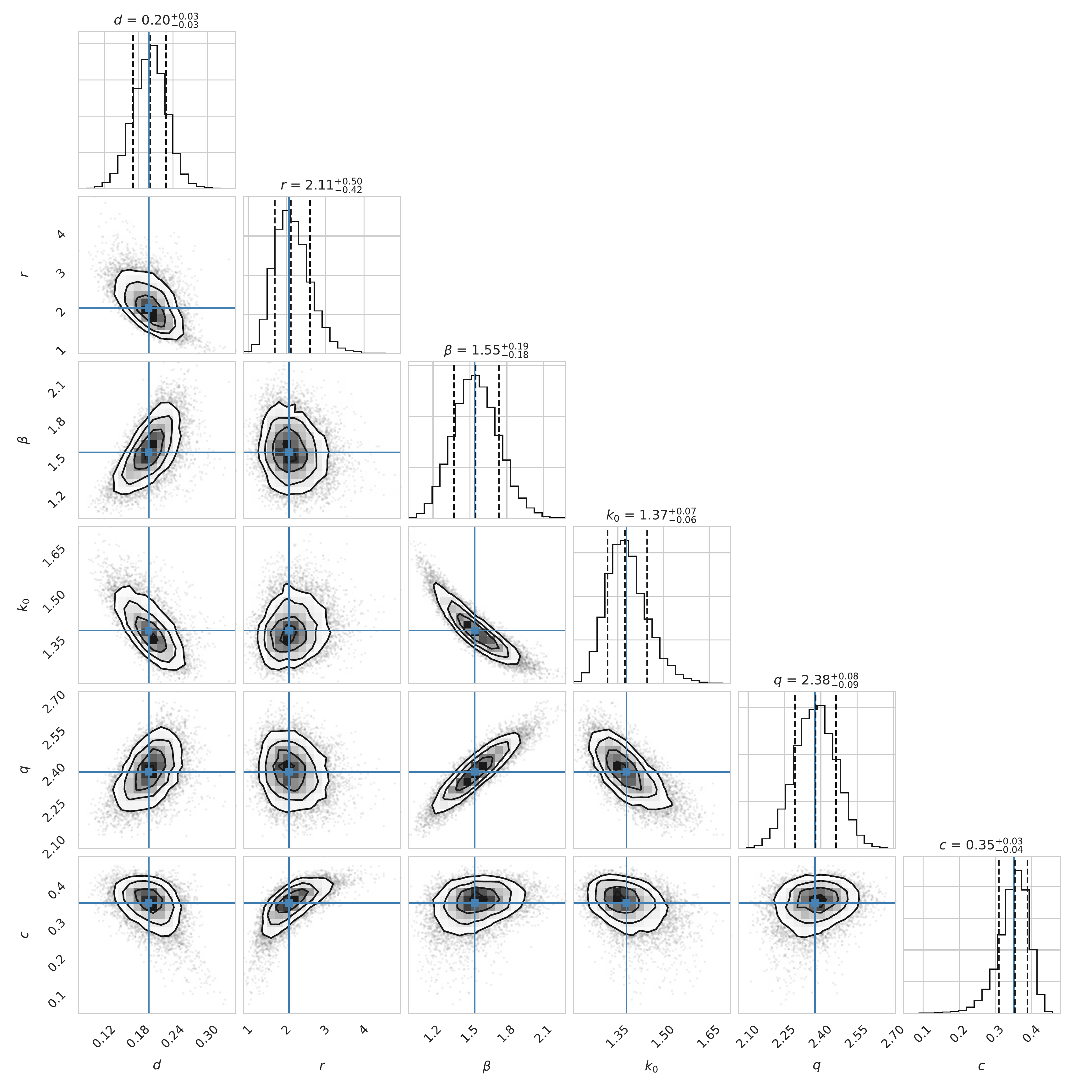}
\caption{The corner plots for the posterior distributions in the Hot (standard case). Above the 1D histogram for each variable we report the 50th quantile and use the 16th and 84th quantiles to compute and show uncertainties. These three quantiles are indicated by the dashed black lines. Contour plots between pairs of parameters are also shown. The blue lines (and dots) represent the value found via the bootstraping procedure. \label{fig4}}
\end{figure*}

\begin{figure*}
\includegraphics[width=\textwidth]{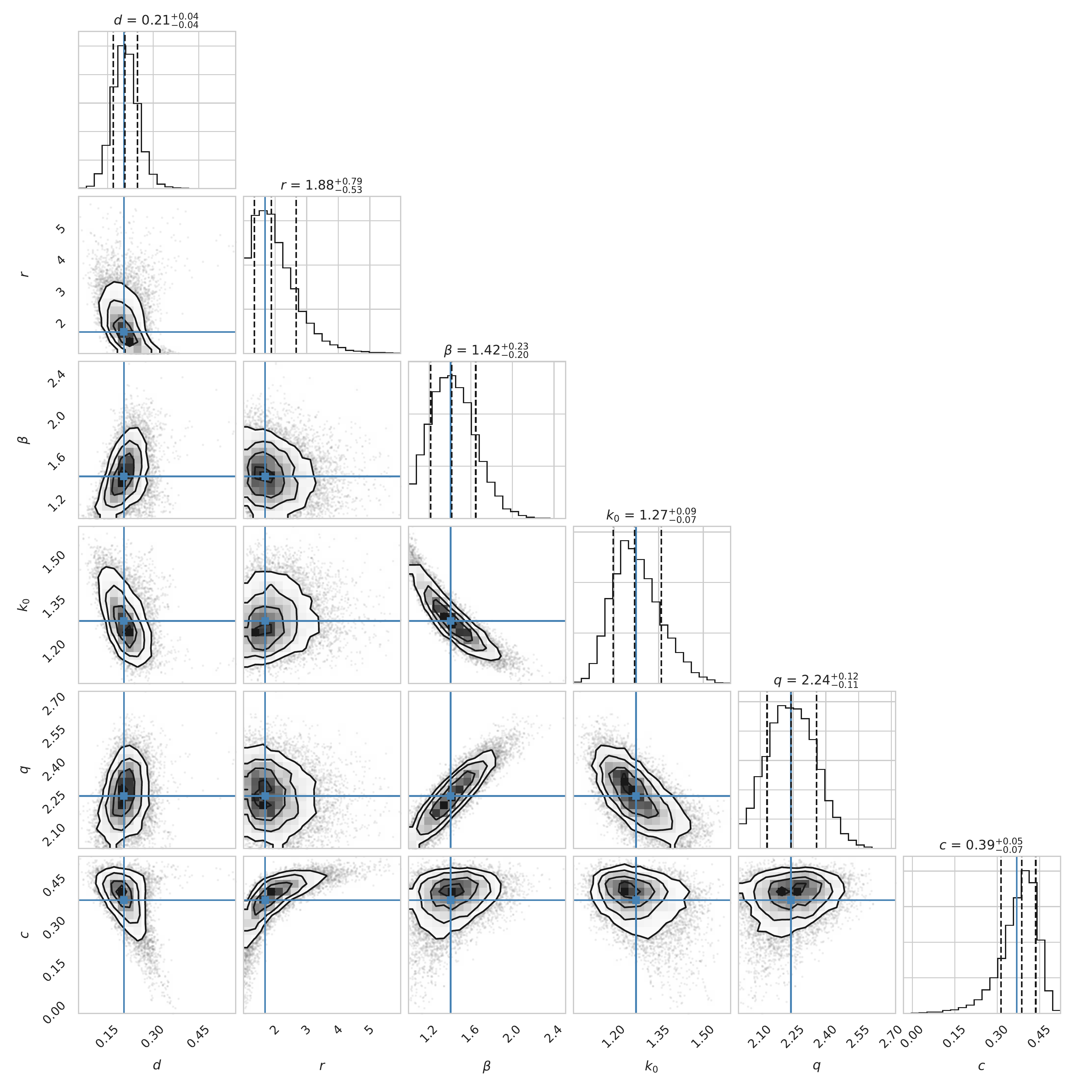}
\caption{Same as Fig. \protect\ref{fig4}, for the Warm case.\label{fig5}}
\end{figure*}

\begin{figure*}
\includegraphics[width=\textwidth]{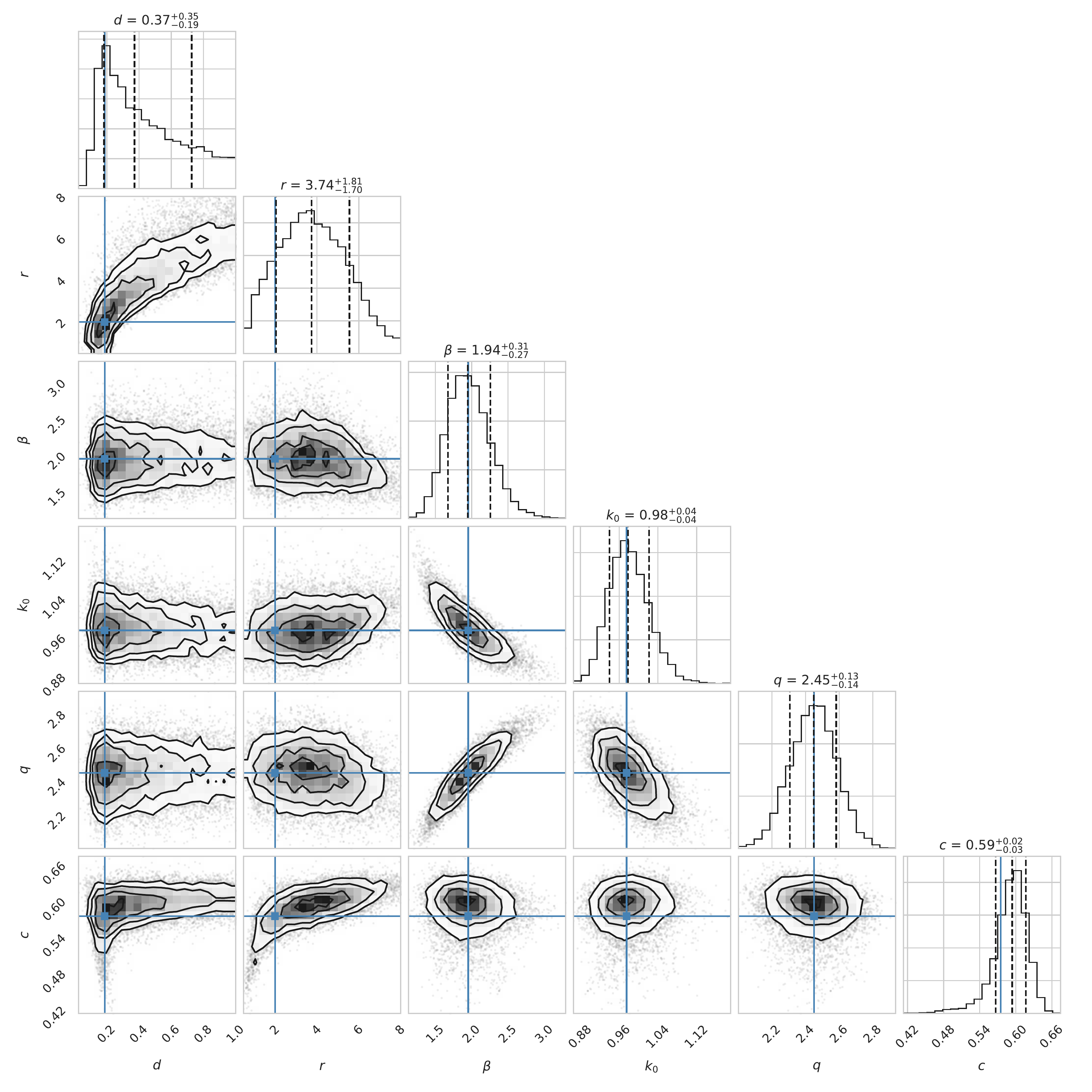}
\caption{Same as Fig. \protect\ref{fig4}, for the Cold case.\label{fig6}}
\end{figure*}

The results of the MCMC analysis are shown in Figures \ref{fig4}, \ref{fig5} and \ref{fig6}, respectively for the Hot, Warm and Cold cases, and for convenience are also summarized in Table V. A first observation is that the minima found by the earlier analysis roughly coincide with the peak in the likelihood given by the MCMC method. The parameter where the largest difference can occur is $r$ where the distribution of the posterior tends to widen significantly as we move to the more cooled cases. Indeed in the Cold case the minimum found in the simpler analysis is not the global minimum. Another interesting change is the behavior of the posterior of $d$ which seems to become more asymmetric---this is evident in figure \ref{fig6}. Note that we quote always the 50th quantile with the 16th and 84th ones being used for the uncertainty calculation both in the aforementioned figures and in Table V, however for parameter $d$ in the Cold case the 16th quantile corresponds to the peak on the posterior distribution (and not the 50th).

Several degeneracies between parameters are worthy of note. For example $k_0$ is negatively correlated with $d$, $\beta$ and $k_0$, while the latter three are positively correlated with one another.  On the other hand, $c$ and $r$ are also positively correlated, and the latter is clearly (but also unsurprisingly) the least well determined VOS parameter. The above correlations can be physically understood in the VOS context. As an example, take the one between $k_0$ and $q$. While $k_0$ is a parameter largely determined by the large expansion rate regime (given that $k(v)$ reduces to it for very low velocities) it also indicates the normalization of the curvature parameter. This normalization, if above unity, is an indication of wiggliness and small scale structure and the point where these features are more obvious is at low expansion rate, having an effect on the mean velocity of strings, and thus on $q$.

Overall we note that the calibrated model parameters are again in good agreement with the previous results, but this agreement worsens with the amount of cooling. The reason for this latter is clear: the VOS model includes terms that separately describe (at least in a statistical sense) energy losses from loop production and radiation. By introducing cooling one forcibly removes some of this radiation from the simulation box, and thereby erases information that is useful for the model calibration.

As for the uncertainties in the VOS parameters, the main result is that the bootstrap methods underestimate the uncertainties in $\beta$ and $r$, and to a lesser extent the uncertainty in $q$. On the other hand, the uncertainties in $c$, $d$ and $k_0$ are only marginally increased in the MCMC case (with the exception of $d$ in the Cold case).

\begin{table*}
\label{table5}
\begin{tabular}{| c | c  c  c  c  c  c | c |}
\hline
Case & d & r & $\beta$ & $k_0$ & q & c & Reference\\
\hline
Hot & $0.20^{+0.03}_{-0.03}$ & $2.11^{+0.50}_{-0.42}$ & $1.55^{+0.19}_{-0.18}$ & $1.37^{+0.07}_{-0.06}$ & $2.38^{+0.08}_{-0.09}$ & $0.35^{+0.03}_{-0.04}$ & This work \\
Warm & $0.21^{+0.04}_{-0.04}$ & $1.88^{+0.79}_{-0.53}$ & $1.42^{+0.23}_{-0.20}$ & $1.27^{+0.09}_{-0.07}$ & $2.24^{+0.12}_{-0.11}$ & $0.39^{+0.05}_{-0.07}$ & This work \\
Cold & $0.37^{+0.35}_{-0.19}$ & $3.74^{+1.81}_{-1.70}$ & $1.94^{+0.31}_{-0.27}$ & $0.98^{+0.04}_{-0.04}$ & $2.45^{+0.13}_{-0.14}$ & $0.59^{+0.02}_{-0.03}$ & This work \\
\hline
\end{tabular}
\caption{Same as Tables III and IV, but using the Bayesien inference method described in the text. We always report the 50th quantile value, with the 16th and 84th being used for computing the uncertainties.}
\end{table*}

\section{\label{conc}Conclusions}

We have used our fast GPU-accelerated Abelian-Higgs string evolution code to quantify the effect of cooled initial conditions, which have been used by several previous authors, on the evolution of the string networks. As a diagnostic in this analysis we have used the values inferred when simulations with different amounts of cooling are used to calibrate the canonical and quantitative VOS model for string network evolution.

Our analysis shows that a modest amount of cooling will have no statistically significant impact on the VOS model calibration, but a stronger (or, in practice, longer) dissipation period does have a noticeable effect. Physically this result is not surprising, but from the point of view of the VOS model itself it also confirms the analysis in \cite{Correia:2019bdl}, in the sense that the model can indeed separate (in a statistical sense) energy losses due to loop production and radiation, since these have different velocity dependencies. These velocity dependencies can therefore be identified (or, perhaps more accurately, reconstructed) by simulating cosmic string networks with many different expansion rates, since the expansion rate will obviously impact the string network velocities.

The logical conclusion is that if the main purpose of simulations is to reach scaling as fast as possible then an early period of cooling is useful, but if the main purpose is to accurately calibrate an analytic model---or indeed study its energy loss mechanisms---then a cooling period is detrimental: one may have a slight gain in the fraction of the simulation time in which the network has reached scaling (as measured by the behaviour of the mean string separation, though not necessarily by that of the average velocity), but this gain is negligible when compared to the loss of information on the radiation in the box, which in practical terms helps to reduce the degeneracies between the model parameters.

We have also taken this opportunity to test and improve the robustness of our VOS calibration pipeline, specifically by implementing a new MCMC based pipeline. A comparison of the results obtained with this pipeline to those from the pipeline used in our previous work (which relied on simpler bootstrap methods) shows that the best-fit values of the VOS model parameters are accurately determined and agree in both pipelines (especially in the Hot case where no cooling is applied), although the previous pipeline did underestimate the uncertainties in some of the VOS parameters. It is also reassuring that the three model parameters whose uncertainties are in better agreement in both pipelines are $c$, $d$ and $k_0$, which happen to be the ones with a more direct physical interpretation (while the other three are more phenomenological). The MCMC analysis is also useful for identifying the degeneracies between the various model parameters. This is useful for planning future sets of simulations, since the constraining power of simulations on various model parameters depends not only on the volumes (in other words, box sizes) that can be simulated but also on the expansion rates being simulated.

An interesting question related to the relative contributions of energy losses from loop production and radiation towards scaling, and how the VOS model may describe both. Clearly in Nambu-Goto simulations only the former is relevant, while in field theory simulations both may contribute, and their relative importance may depend on various physical and numerical parameters. From a purely mathematical perspective the model can clearly describe both regimes (and scaling will be an attractor in either case), although it is not a prior clear, for example, if the same model parameters should apply to Nambu-Goto and Abelian-Higgs calibrated models: it is conceivable that some parameters remain unchanged while others do change.

Numerically, thus far we have used simulation boxes that are relatively small and might conceivably not possess the dynamic range necessary to achieve scaling sustained by massive radiation alone. This is one possible interpretation of the results of  \cite{Klaer:2019fxc} where $2048^3$ simulation boxes achieved smaller values of the loop chopping parameter than those that we report in this work, though it should also be noted that the analysis method therein is substantially different from ours. Larger simulation boxes (allowing an increased dynamic range for scaling) and a more extensive exploration of the space of relevant numerical simulation parameters will be necessary to fully assess the robustness of the model while improving our understanding of the underlying physics of scaling.

Overall, our analysis shows that the calibration pipeline is robust and can be applied to much larger field theory simulations, which are enabled by our highly efficient GPU-accelerated code. In particular, work on $8192^3$ box simulations, which can now be done in high-performance computing facilities, is in progress \cite{mGPU}. We expect to report on the results of these in the near future.

\begin{acknowledgments}
This work was financed by FEDER---Fundo Europeu de Desenvolvimento Regional funds through the COMPETE 2020---Operational Programme for Competitiveness and Internationalisation (POCI), and by Portuguese funds through FCT - Funda\c c\~ao para a Ci\^encia e a Tecnologia in the framework of the project POCI-01-0145-FEDER-028987 and PTDC/FIS-AST/28987/2017. J.R.C. is supported by an FCT fellowship (SFRH/BD/130445/2017). We gratefully acknowledge the support of NVIDIA Corporation with the donation of the Quadro P5000 GPU used for this research. J.R.C. acknowledges Jo\~ao Faria's Programmer's Club presentation for introducing him to MCMC and emcee usage. J.R.C. also acknowledges Luisa Maria Serrano's and Jo\~ao Camacho's assistance with MCMC and general emcee usage.
\end{acknowledgments}
 
\bibliography{artigo}
\end{document}